\documentclass[final,5p,times,twocolumn]{elsarticle}

\usepackage{graphicx}
\usepackage{subfigure}
\usepackage{multirow}
\usepackage{wrapfig}
\usepackage{url}
\usepackage{dblfloatfix}
\usepackage{float}
\usepackage{tabularx}
\usepackage{booktabs}
\usepackage{xcolor}
\usepackage[utf8]{inputenc}
\usepackage{amsmath}
\usepackage{amsthm}
\usepackage{amsfonts}
\usepackage{epsfig}
\usepackage{psfrag}
\usepackage{pstricks}
\usepackage{algorithm}

\usepackage{algorithmicx}
\usepackage{algpseudocode}
\floatname{algorithm}{Algorithm}

\graphicspath{{./Figures/}}

\usepackage[normalem]{ulem}
\usepackage{cancel}

\usepackage{amssymb}
\usepackage{bm}

\biboptions{comma,sort&compress}

\usepackage{enumitem}
\setlist{nosep,topsep=-\parskip}

\usepackage{color}
\usepackage{ulem}

\usepackage{xcolor}
\usepackage{tikz}

\journal{Under Review}
\usepackage[export]{adjustbox}
\usepackage{array}

\begin{document}
\baselineskip11pt

\begin{frontmatter}

\title{KDH-CAD: Knowledge–data hybrid CAD learning under data scarcity}

\author[]{Ziqin Gao}
\author[]{Zhijie Yang}
\author[]{Qiang Zou\corref{cor}}\ead{qiangzou@cad.zju.edu.cn}

\cortext[cor]{Corresponding author.}
\address{State Key Laboratory of CAD$\&$CG, Zhejiang University, Hangzhou, 310027, China}

\begin{abstract}
Deep learning in computer-aided design (CAD) remains fundamentally constrained by the data scarcity challenge: authentic CAD data is difficult to collect at scale, while synthetic data may not faithfully reflect real design practice. Rather than pursuing ever-larger CAD datasets, this paper alternatively treats CAD learning as a knowledge completion and calibration problem. It introduces KDH-CAD, a knowledge–data hybrid framework that integrates pretrained knowledge in foundation models, structured domain knowledge from textbooks/tutorials, and a very small amount of labeled CAD data. Domain knowledge is used to elicit and complete CAD-relevant concepts that are weakly expressed or under-represented in pretrained foundation models, while labeled CAD data calibrates these concepts in the latent space to account for task-specific geometric variability, without fine-tuning the foundation model. Experiments on real-world mechanical part classification show that KDH-CAD achieves strong performance in low-data regimes, reaching 92.6\% accuracy with only 250 training samples, 95.8\% with 1,000 samples, and continuing to improve with additional data. This matches or exceeds state-of-the-art performance that typically requires an order of magnitude more data. These results suggest that combining pretrained foundation models with structured domain knowledge can substantially reduce reliance on large-scale CAD datasets, providing a principled and practical direction for data-efficient CAD learning.
\end{abstract}


\end{frontmatter}

\section{Introduction}
\label{sec:intro}

Computer-aided design (CAD) is fundamental to modern engineering workflows, enabling design reuse, analysis, and manufacturing planning~\cite{zou2023variational}. Learning-based methods are increasingly explored to bring intelligence into CAD, with the potential to improve productivity and facilitate design space exploration~\cite{2021_Model_Retrieval_AI_based_UVNet}. A key challenge to this integration is, however, data scarcity~\cite{zou2024intelligent}. Unlike natural language and images, real-world CAD models are difficult to collect at scale: industrial CAD data are typically proprietary and require substantial expertise and modeling effort. As a result, building large and high-quality CAD datasets is costly, which is one of the primary reasons why learning-based CAD methods have lagged behind other data-rich fields.

To mitigate this limitation, a common strategy is to scale CAD data synthetically, for example, by procedural generation or parametric variation, with datasets such as ABC~\cite{koch2019abc}, SketchGraphs~\cite{SketchGraphs}, and CADL~\cite{ganin2021computer} being typical efforts. Synthetic datasets have enabled meaningful progress and become important resources for benchmarking. However, they may differ from real-world design practice in important ways, such as feature history decisions and design intent, because feature histories often reflect functional, manufacturing, and modification requirements, while design intent captures geometric constraints and feature dependencies. In contrast, synthetic CAD data are often generated by rule-based or randomized procedures, so their histories and intent may not correspond to realistic engineering decisions, which can limit their applicability to practical design scenarios~\cite{2026_zhou_CADialogue}.

In this work, we explore an alternative direction: rather than pursuing ever-larger CAD datasets, we aim to leverage pretrained knowledge in foundation models, specifically vision–language models (VLMs), to reduce reliance on large amounts of CAD data. Our motivation is that, while foundation models are trained on broad corpora where CAD is not a primary focus, they may still encode partial CAD-relevant concepts (e.g., part type and function-related concepts). If these concepts can be elicited and completed, the integration of AI and CAD can focus on knowledge calibration under limited CAD data, instead of exhaustive data scaling.

While appealing, realizing this idea is not trivial due to the following reasons:
\begin{itemize}
    \item \textbf{Entanglement with general knowledge}:
    In generic pretraining corpora, CAD-relevant material is mixed with broad visual and linguistic content. CAD-relevant concepts could thus be weakly expressed and even suppressed by dominant general concepts. Consistent with this, our experiments show that directly applying a pretrained VLM (for example, Qwen3-VL-2B-Instruct~\cite{Qwen3-VL}) to mechanical part classification yields only $\sim$84\% accuracy (Sec.~\ref{sec:result}), far below what is required for practical use.

    \item \textbf{Incompleteness of CAD-relevant concepts}:
    CAD-relevant material constitutes a sparse and imbalanced fraction of pretraining corpora. Moreover, even within available CAD-relevant material, mechanical part categories typically follow long-tail frequency distributions such that several functionally important parts (e.g., gaskets) appear rarely (see Chapter 3 in~\cite{fsa_esa_gasket_2017}). Therefore, CAD-relevant concepts can be under-represented or even missing in foundation models, leading to ambiguity in downstream tasks.

    \item \textbf{Concept-geometry mismatch}:
    CAD-relevant concepts are tied to design intent and function rather than geometry alone (e,g. NURBS~\cite{zOU2025splinegen}). Within a single part category, geometry may vary substantially (e.g., brackets). This concept-geometry mismatch limits purely vision-grounded foundation models and reduces robustness in downstream applications.
\end{itemize}

To address these challenges, we introduce KDH-CAD, a knowledge–data hybrid learning approach that integrates pretrained foundation models, structured domain knowledge, and small data in a unified and complementary manner. The key principle is that foundation models provide a transferable representation base, domain knowledge provides anchors for CAD-relevant concepts, and a small amount of labeled data calibrates these concepts to account for geometric variability in the target task. Particularly, we use domain knowledge to extract reusable latent embeddings that represent CAD-relevant concepts and calibrate these embeddings with small data toward the target task while keeping the foundation model frozen. This contrasts with prompting/RAG approaches~\cite{2026_zhou_CADialogue, 2025_Kamel_VLM_CAD_Code_Generation, 10.1115/DETC2024-143740, 10890248} that incorporate knowledge by inference-time text conditioning, and with fine-tuning methods~\cite{zhang2024flexcad, yuan2025cad, wang2025texttocad, mews2025dontmeshmegenerating, llm4cad} that update model parameters using task data. A more detailed discussion is provided in Sec.~\ref{sec:foundationmodel}.

Specifically, KDH-CAD leverages mechanical design textbooks and tutorial videos to elicit and complete CAD-relevant concepts in pretrained foundation models. These sources provide authoritative domain knowledge, including canonical terminology, definitions, and illustrative examples. Conditioning foundation models on this knowledge biases them toward CAD-relevant concepts during embedding extraction. This has two effects: (1) it elicits CAD-relevant concepts that are otherwise suppressed by general concepts, and (2) it expands concept coverage by supplying explicit anchors for rare but important part types.

While domain knowledge strengthens CAD-relevant concepts, it is largely task- and dataset-independent and cannot by itself solve the concept-geometry mismatch problem. KDH-CAD addresses this with a lightweight knowledge calibration network that learns additive shift vectors in the foundation model's embedding space using a small labeled CAD dataset. Intuitively, the shift vectors re-center knowledge embeddings to better match the target data distribution and account for geometric variations under the
same concept.

Overall, KDH-CAD uses domain knowledge to establish what CAD-relevant concepts should be, and limited data to determine how these concepts adapt to geometry variations in a specific task setting. This mirrors a typical engineering student learning process: concepts are acquired from authoritative references (textbooks or lectures) and then calibrated through a small set of representative exercises. Note that, although knowledge–data hybrid learning is intended as a general method under data scarcity, its implementation is task-dependent because CAD tasks differ in inputs, objectives, and forms of domain knowledge. In this work,  we demonstrate the effectiveness of this idea on part classification, a fundamental capability that can be integrated into CAD/CAM workflows as an auxiliary module for downstream applications such as design reuse and retrieval, BOM/assembly reasoning, manufacturing analysis, and annotation. Part classification also offers established baselines and widely used benchmarks~\cite{2021_Model_Retrieval_AI_based_UVNet, 2024_Wu_AAGNet, zou2025bringing}, enabling controlled evaluation.

This paper's contributions can be summarized as follows:
\begin{itemize}

    \item We introduce KDH-CAD, a knowledge–data hybrid framework that integrates pretrained foundation models, structured domain knowledge, and a very small amount of labeled CAD data to solve the issue of data scarcity in CAD learning.

    \item We propose a new part classification method based on the KDH-CAD framework and achieve state-of-the-art performance with only hundreds of training samples, while existing methods require tens of thousands.

    \item We present an alternative way to data scaling and empirically demonstrate that it is a promising and practical solution for data-efficient CAD learning.
\end{itemize}

The remainder of this paper is organized as follows: Sec.~\ref{sec:related work} provides a review of related studies. Sec.~\ref{sec:method} elaborates the proposed KDH-CAD approach. Validation of the method using a series of examples and comparisons can be found in Sec.~\ref{sec:result}, followed by conclusions in Sec.~\ref{sec:conclusion}.

\section{Related work}
\label{sec:related work}

In this section, we review prior efforts along three axes that directly relate to our work: CAD datasets, CAD learning with foundation models and limited data, and part classification.

\subsection{CAD Datasets}
\label{sec:CADdataset}

Large-scale datasets have played an important role in recent advances in language and vision, which has motivated growing interest in constructing datasets for CAD learning. For solid modeling, several datasets have been developed based on boundary representations (B-rep)~\cite{zou2019push}, including widely used benchmarks such as ABC~\cite{koch2019abc} and MFCAD++~\cite{colligan2022hierarchical}, along with subsequent extensions and variants~\cite{cao2020graph, lambourne2021brepnet, jayaramansolidgen, zhang2024brepmfr, cheng2025constraint, zou2025bringing}. Beyond solid models, parametric CAD datasets have been proposed to capture design histories, features, and constraints~\cite{tang2023decision}, with representative examples including SketchGraphs~\cite{SketchGraphs}, HPSketch~\cite{fan2025history}, DeepCAD~\cite{wu2021deepcad}, and the Fusion 360 gallery~\cite{willis2021fusion}. In addition, datasets targeting higher-level structural organization, such as assembly modeling, have been introduced, including AutoMate~\cite{jones2021automate} and JoinABLe~\cite{willis2022joinable}. Collectively, these datasets have become important sources for evaluating learning-based methods in CAD-specific tasks, such as sketch generation, part classification, feature recognition, and mating prediction.

However, many existing CAD datasets remain limited in scale, often containing only hundreds of thousands of samples, which may constrain the generalization of CAD learning methods. Only a very small number of datasets (e.g., ABC~\cite{koch2019abc}, SketchGraphs~\cite{SketchGraphs}, and CADL~\cite{ganin2021computer}) are large-scale ones, reaching the million level. Moreover, these datasets are synthetic, which constructs the models by randomly modifying parametric models or by collecting models from novice users. Because they are not curated from real-world engineering CAD workflows, their practical relevance to design practices remains unknown.

\subsection{CAD Learning with Foundation Models and Limited Data}
\label{sec:foundationmodel}

Our work involves leveraging foundation models and limited data, which is naturally connected to Retrieval Augmented Generation (RAG), Prompt Engineering, Supervised Finetuning on foundation models, and Few-shot Learning.

\textbf{Prompt engineering and RAG.} Prompt engineering and RAG improve foundation model performance by manipulating the input space with designed prompts or retrieved context, without updating any parameters. For instance, methods such as~\cite{2026_zhou_CADialogue, 2025_Kamel_VLM_CAD_Code_Generation, 10.1115/DETC2024-143740} design structured prompts to guide foundation models toward modeling sequence generation, and ChatCAD~\cite{10890248} retrieves relevant CAD context to refine model responses. These methods require carefully designed prompts to guide model outputs, whereas we operate directly on latent embeddings, avoiding such manual design.

\textbf{Foundation model fine-tuning.} Another line of work modifies the parameters of foundation models through supervised fine-tuning or reinforcement learning. In the CAD domain, methods such as~\cite{zhang2024flexcad, yuan2025cad, wang2025texttocad, mews2025dontmeshmegenerating, llm4cad,yuan20243d} adopt these fine-tuning strategies to adjust the foundation model for desired CAD model generation. However, fine-tuning is challenging due to the large number of free parameters, which can lead to instability and high computational cost. In contrast, our approach keeps the foundation model frozen, ensuring stable and efficient training process.

\textbf{Few-shot learning.} To address the challenge of data scarcity, few-shot learning has emerged as a primary approach. Within few-shot learning, we focus on metric-learning methods, as they are one of the few approaches explored in the CAD domain. Prototypical networks~\cite{snell2017prototypical} are among the well-known methods for initializing metric-learning models, and in the CAD domain, Pcd-Prototype~\cite{ZHOU2026103976} applies this idea to segment point clouds. Although effective, such methods rely solely on the information available in the data and do not incorporate domain knowledge. Our method can also be viewed as a form of few-shot learning, but differs in its core mechanism: rather than adapting parameters solely from limited data, our approach grounds learning on CAD priors.

\subsection{Part Classification}
\label{sec:partclassification}

Part classification is the task chosen to demonstrate the effectiveness of the proposed method. To carry out this task, traditionally, rule-based methods are adopted. Specifically, these methods extract geometric and topological features from the CAD models for similarity matching~\cite{2009_Li_Model_Retrieval_Traditional_JMD, 2015_Model_Retrieval_Traditional_CAD}, but those features are constructed in a heuristic way. Therefore, they are not comprehensive and are limited to specific application scenarios.

Recent works have shifted to a data-driven paradigm, with development broadly falling into two categories: conversion-based and direct methods. Conversion-based methods first transform CAD models into intermediate schemes (e.g., voxels, point clouds, meshes, or multi-view images), and then use the well-established deep learning networks from computer vision and graphics to carry out the classification task. Notable examples include Wang et al.~\cite{2017_Wang_OCNN} for voxels, Bickel et al.~\cite{2023_Bickel_Model_Retrieval_AI_based_Object_Projection} for point clouds, Liang et al.~\cite{2022_Liang_MeshMAE} for meshes, and Zhang et al.~\cite{2020_Zhang_Model_Retrieval_AI_based_View_based} and G\"{u}meli et al.~\cite{2022_Gumeli_Model_Retrieval_AI_based_ROCA} for view-based images. In contrast, direct methods learn from the geometric and topological data in CAD models. The main idea~\cite{2021_Model_Retrieval_AI_based_UVNet, 2023_Hou_Model_Retrieval_AI_based_FusGCN, 2025_Qin_Model_Retrieval_AI_based_CADGCL, 2023_Lou_Model_Retrieval_AI_based_BrepBert} is to represent CAD models as face-adjacency graphs, and apply graph neural networks for part classification. More recently, transformer-based architectures have also been introduced, achieving better results~\cite{zou2025bringing}.

To date, substantial progress has been made using specialized models to carry out CAD tasks, such as part classification. Nevertheless, CAD learning under data scarcity has yet been made available. This work explores an alternative to solve this issue.

\begin{figure*}[t]
  \centering
  \includegraphics[width=0.8\textwidth]{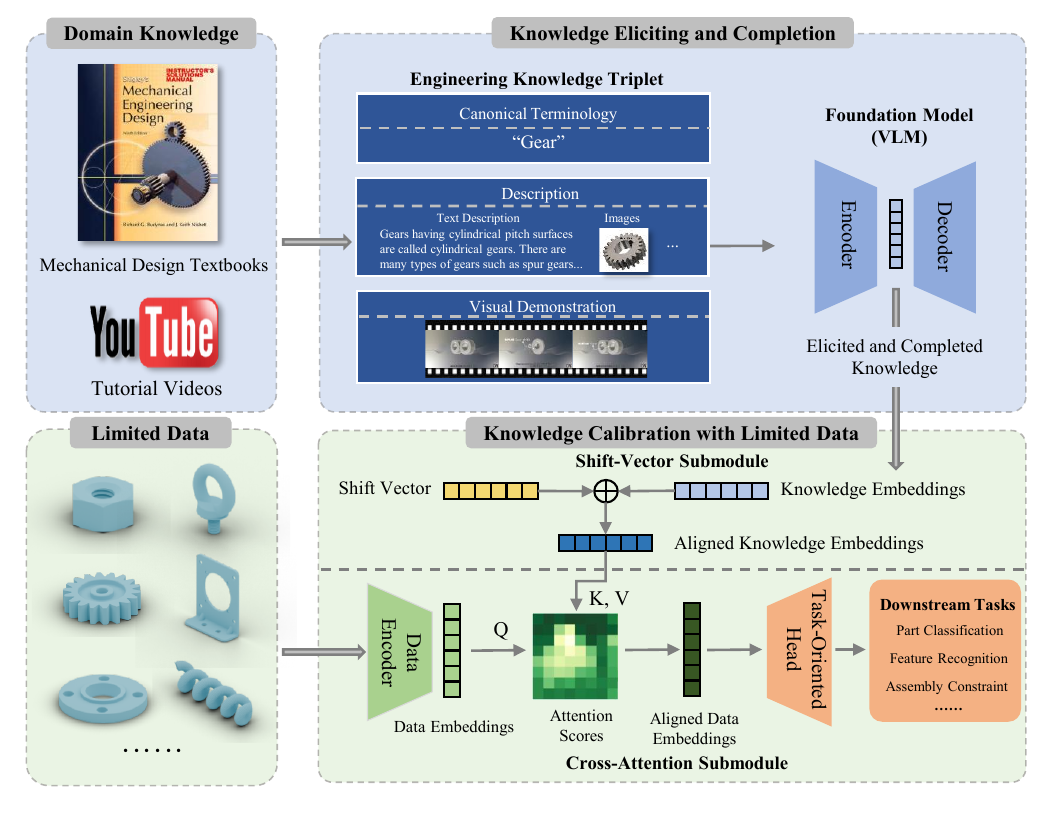}
  \caption{The overall architecture of our general KDH-CAD.}
  \label{fig:generalArchitecture}
\end{figure*}

\noindent

\section{Methods}
\label{sec:method}

KDH-CAD is designed for CAD learning under data scarcity by utilizing pretrained knowledge in foundation models. It decomposes the learning process into two stages, as illustrated in Fig.~\ref{fig:generalArchitecture}. First, we leverage domain knowledge (in textbooks and tutorial videos) to elicit and complete CAD-relevant concepts within a pretrained foundation model, with details in Sec.~\ref{sec:method-knowledge_elicting_and_completion}. Second, we calibrate such CAD-relevant concepts to the target data distribution in a specific task via a knowledge calibration network to solve the concept-geometry mismatch issue, which is described in Sec.~\ref{sec:method-aligning_knowledge_with_limited_data}. We then instantiate and demonstrate the proposed method in the context of part classification, with the specialized design presented in Sec.~\ref{sec:method-specialized_design_for_part_classification}.

\subsection{Knowledge Eliciting and Completion}
\label{sec:method-knowledge_elicting_and_completion}

As discussed in the introduction section, CAD-relevant concepts can be weakly expressed or under-represented in foundation models. To address this issue, KDH-CAD utilizes domain knowledge to elicit and complete CAD-relevant concepts in the foundation model, specifically a VLM (in principle, any VLM can be used; Qwen3-VL-2B-Instruct~\cite{Qwen3-VL} is adopted in this work for its strong multimodal and instruction-following ability). More concretely, mechanical design textbooks and tutorial videos are chosen as the domain knowledge because (1) textbooks provide engineering knowledge with standard terminology and definitions, and (2) tutorial videos, often presented in an animated and step-by-step form, offer intuitive demonstrations that facilitate comprehension. During triplet construction, the VLM is used only to convert these materials into a structured triplet format.

Textbooks and videos were initially directly fed into the foundation model, but this straightforward strategy does not yield good results as reported in the ablation studies in Sec.~\ref{sec:result}. This is largely because they are not presented in a structured way to facilitate the VLM's understanding of domain knowledge. Instead, this work organizes textbooks and tutorial videos into a structured representation, consisting of:
\begin{itemize}

    \item \textbf{Canonical terminology}:
    A word (or short phrase) that uniquely identifies the target CAD semantics.

    \item \textbf{Description}:
    A textual description that summarizes the definition, primary function, typical variants, and common usage scenarios, paired with illustrative images.

    \item \textbf{Visual demonstration}:
    A tutorial video that provides a step-by-step visual explanation, highlighting representative appearances and typical usage in practice.

\end{itemize}

For simplicity, we call this representation an engineering knowledge triplet hereafter as located upper in Fig.~\ref{fig:generalArchitecture}. The three components of the triplet play complementary roles. Canonical terminology standardizes naming by grouping paraphrases and synonyms under a consistent formulation, which reduces ambiguity and clarifies concept boundaries. Description distills the key meaning of the concept, avoiding redundant textbook details while retaining information that is useful for discrimination. Visual demonstration provides a consistent visual reference, allowing the VLM to associate the terminology and description with visual content. More broadly, this triplet is defined at the level of concepts rather than any specific task. Therefore, it provides a general representation for different learning tasks (excluding generation).

\setlength{\intextsep}{2pt}
\begin{table}[!h]
    \centering
    \small
    \setlength{\tabcolsep}{4pt}
    \renewcommand{\arraystretch}{1.25}
    \caption{Example of the engineering knowledge triplet for a gear.}
    \begin{tabular}{m{0.2\linewidth} m{0.65\linewidth}}
        \toprule
        \textbf{Component} & \textbf{Content} \\
        \midrule
        Canonical Terminology &
        Gear \\

        Description &
        {\footnotesize
        A gear is a kind of machine element in which ... they are used to transmit rotations and forces ... There are many types of gears such as spur gears, helical gears, ... These can be broadly classified by looking at the positions of axes ...
        (Images: Illustrative images of the gear associated with the above text)
        } \\

        \mbox{Visual De-} \mbox{monstration} &
        {\footnotesize
        (Video: a tutorial video of the gear)
        } \\
        \bottomrule
    \end{tabular}
    \label{tab:ekt_example_gear}
\end{table}

Take the gear as an example, its engineering knowledge triplet is represented in Table~\ref{tab:ekt_example_gear}. This example demonstrates how canonical terminology, description, and visual demonstration are structured together in our representation.

We then combine the engineering knowledge triplet with the VLM to obtain the elicited and completed knowledge (right and upper side of Fig.~\ref{fig:generalArchitecture}). In the VLM, pretrained knowledge is represented as vector embeddings in its latent space. Therefore, we encode the engineering knowledge triplet in the same latent space to obtain the knowledge embedding that captures elicited and completed CAD-relevant concepts.
This fusion is mutually beneficial: domain knowledge sharpens concept boundaries and facilitates the VLM to elicit and complete pretrained knowledge, while the pretrained knowledge converts the domain knowledge into a compact and consistent representation in the VLM latent space. Consequently, this knowledge embedding provides a transferable representation of CAD-relevant concepts that can be readily utilized in downstream learning tasks.

\subsection{Knowledge Calibration with Limited Data}
\label{sec:method-aligning_knowledge_with_limited_data}

Having obtained knowledge embeddings, they should be calibrated with the target data to solve the concept-geometry mismatch problem. Specifically, a knowledge calibration network is introduced, as illustrated in the lower part of Fig.~\ref{fig:generalArchitecture}. It comprises three components: (1) a geometry-aware data encoder, (2) a knowledge–data alignment module, and (3) a task-oriented head. Specifically, the geometry-aware data encoder extracts task-specific data embeddings from CAD inputs; the knowledge–data alignment module adapts the knowledge embedding to the target data distribution and aligns it with the data embedding; and the task-oriented head maps the aligned representation to the final task prediction. Typically, the data encoder can reuse the same encoder of the foundation model, and the task-oriented head can be implemented as an MLP. This paper designs the knowledge–data alignment module with two submodules: the shift-vector submodule and the cross-attention submodule.

\begin{figure*}[t]
  \centering
  \includegraphics[width=\textwidth]{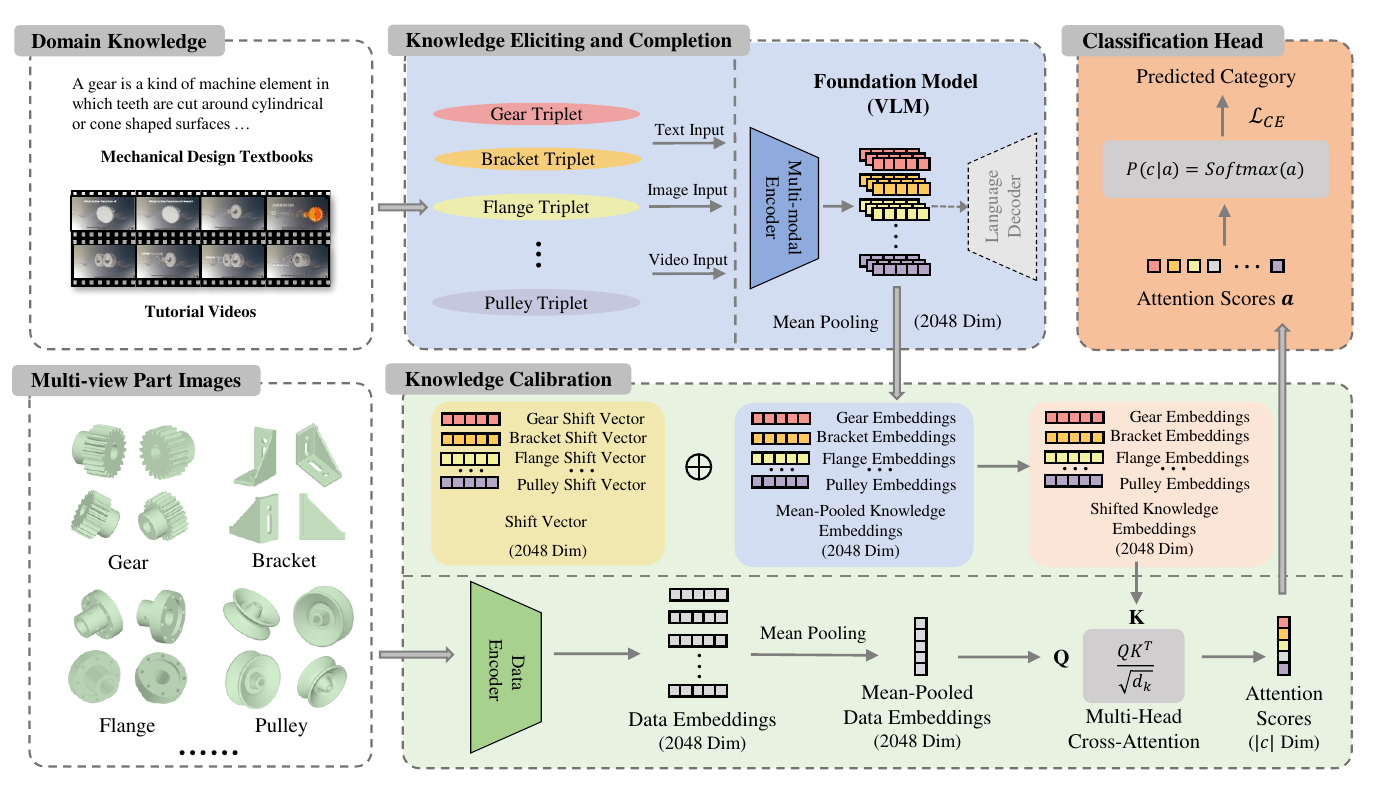}
  \caption{The overall architecture of specialized KDH-CAD for part classification.}
  \label{fig:specializedArchitecture}
\end{figure*}
\textbf{Shift-vector submodule.} To adapt knowledge embeddings toward the target data distribution, we introduce a shift vector to capture the discrepancy between the elicited and completed knowledge and its dataset-adapted counterpart (see shift-vector submodule in Fig.~\ref{fig:generalArchitecture}). Conceptually, for CAD-relevant concepts, the shift vector can be interpreted as the offset between the original knowledge embedding and the dataset-adapted knowledge embedding. Such discrepancies vary across datasets, feature spaces, and data distributions, which require dataset- and method-adapted knowledge embeddings. For this reason, we model the shift vector as a conditional distribution conditioned on both the current knowledge embedding and task-specific data.

Formally, let $\mathcal{K}=[k_1,\ldots,k_N]^{\top}\in\mathbb{R}^{N\times d}$ denote the knowledge embeddings obtained from the VLM for $N$ CAD-relevant concepts, and let $\mathcal{G}_{\mathcal{D}}=[g_1,\ldots,g_T]^{\top}\in\mathbb{R}^{T\times d}$ denote the task-specific data embeddings extracted from the target dataset by the data encoder. For each concept $i$, the shift vector can be represented as:
\begin{equation}
s_i \sim q_{\phi}\!\left(s_i \mid k_i, \mathcal{G}_{\mathcal{D}}\right),
\quad
\tilde{k}_i = k_i + s_i,
\quad i=1,\ldots,N,
\label{eq:shift_distribution}
\end{equation}
where $q_{\phi}(\cdot)$ denotes the conditional distribution of the shift vector, and $\tilde{k}_i$ is the dataset-adapted knowledge embedding.

\textbf{Cross-attention submodule.} To further align the adapted knowledge embeddings (from using shift vectors) with the data embedding, cross-attention is used (see cross-attention submodule in Fig.~\ref{fig:generalArchitecture}). Intuitively, the adapted knowledge embeddings provide a reference for CAD-relevant concepts, and the data embedding acts as a query over these knowledge embeddings, assigning higher weights to more relevant knowledge embeddings and aggregating them to form a more accurate representation.

Formally, let $\{\mathcal{K}_i\}_{i=1}^{N}$ denote the set of adapted knowledge embeddings with size $N$ and let $\mathcal{G}$ denote the data embedding. Cross-attention first maps these embeddings to the same dimension and computes attention scores between the data embedding and the knowledge embedding set:
\begin{equation}
\begin{aligned}
\mathcal{G}_\mathrm{query} &= \mathcal{G}W_\mathrm{query},\quad
{\mathcal{K}_\mathrm{key}}_i = \mathcal{K}_iW_\mathrm{key} \\
\mathcal{A} \;&=\; \operatorname{Softmax}\!\left(\frac{\mathcal{G}_\mathrm{query}\,\cdot\operatorname{Concat}\left(\{{\mathcal{K}_\mathrm{key}}_i\}_{i=1}^{N}\right)^{\top}}{\sqrt{d}}\right),
\end{aligned}
\label{eq:cross_attn_scores}
\end{equation}
where $W_\mathrm{query}$ and $W_\mathrm{key}$ are all learnable projection matrices with the same projected dimension $d$. The aligned data embedding is then obtained by aggregating the knowledge embedding set weighted by the attention scores:
\begin{equation}
\begin{aligned}
{\mathcal{K}_\mathrm{value}}_i &= \mathcal{K}_iW_\mathrm{value} , \\
\mathcal{G}' \;&=\; \mathcal{A}\,\cdot\operatorname{Concat}\left(\{{\mathcal{K}_\mathrm{value}}_i\}_{i=1}^{N}\right)^{\top}.
\end{aligned}
\label{eq:cross_attn_aggregation}
\end{equation}
where $W_\mathrm{value}$ is a learnable projection matrix with the projected dimension $d$.
As a result, the aligned data embedding is informed by both geometric variations and CAD-relevant concepts, making it more suitable for downstream tasks.

\subsection{Specialized Design for Part Classification}
\label{sec:method-specialized_design_for_part_classification}

To evaluate the effectiveness of the proposed KDH-CAD method, we instantiate it in the context of part classification. As illustrated in Fig.~\ref{fig:specializedArchitecture}, the part classification architecture follows our general pipeline, including knowledge eliciting and completion, and knowledge calibration with limited data, while instantiating both components with classification-specific designs.

For knowledge eliciting and completion, domain knowledge (in mechanical design textbooks and tutorial videos) is curated and formatted into a structured input, then processed by the VLM (i.e., Qwen3-VL-2B-Instruct) to obtain the knowledge embedding. A key difference from a single-pass extraction strategy is that we construct the part-specific knowledge and query the VLM independently for each category, so that the resulting embedding is explicitly anchored to the concept of that category and avoids negative interplay by different concepts. This produces a set of category-level knowledge embeddings that will later guide knowledge calibration.

To utilize the above strategy, we first instantiate the structured input for each part category in the form of the engineering knowledge triplet as described in Sec.~\ref{sec:method-aligning_knowledge_with_limited_data}. Specifically, the triplet is instantiated as a paired text description and an illustrative image (covering canonical terminology and description) together with a tutorial video (as visual demonstration) \footnote{In the following, the text description, illustrative image, and tutorial video are referred to as text, image, and video, respectively.}.

Formally, let $c \in \mathcal{C}$ denote a part category. Given the text, the image, and the video, we use the VLM's language encoder (specifically, a word tokenizer) to encode the text and the VLM's vision encoder to encode the image and the video into tokens:
\begin{equation}
\begin{aligned}
R_c &= f_{\mathrm{LE}}(\textit{text}) \in \mathbb{R}^{{L_c}_\mathrm{text} \times d}, \\
U_c &= f_{\mathrm{VE}}(\textit{image}) \in \mathbb{R}^{{L_c}_\mathrm{image} \times d}, \\
V_c &= f_{\mathrm{VE}}(\textit{video}) \in \mathbb{R}^{{L_c}_\mathrm{video} \times d}.
\end{aligned}
\label{eq:text_image_tokens}
\end{equation}
where $R_c$, $U_c$, and $V_c$ denote the text, image, and video token sequences, with lengths ${L_c}_\mathrm{text}$, ${L_c}_\mathrm{image}$, and ${L_c}_\mathrm{video}$, respectively, and $d$ is the hidden dimension of the VLM ($d=2048$ in our implementation). These token sequences are then converted into a multi-modal token sequence by concatenation:
\begin{equation}
H_c = \operatorname{Concat}(R_c, U_c, V_c) \in \mathbb{R}^{({L_c}_\mathrm{text}+{L_c}_\mathrm{image}+{L_c}_\mathrm{video})\times d}.
\label{eq:multimodal_concat}
\end{equation}

Finally, the token sequence is fed into a VLM encoder $f_\mathrm{ME}$ that maps multi-modal tokens into the latent space, to obtain the token-level knowledge embedding for category $c$:
\begin{equation}
K_c = f_{\mathrm{ME}}\!\left(\operatorname{Concat}(R_c, U_c, V_c)\right) \in \mathbb{R}^{L_c \times d},
\label{eq:token_level_knowledge}
\end{equation}
where $L_c$ ($= {L_c}_\mathrm{text} + {L_c}_\mathrm{image} + {L_c}_\mathrm{video}$) is the length of the resulting knowledge embedding sequence.

For knowledge calibration, we instantiate the knowledge calibration network with three components: (1) a data encoder, (2) a knowledge–data alignment module, and (3) a classification head. To facilitate alignment with multimodal knowledge in the VLM, each part model is rendered into multi-view images as the task-specific representation, and we use the same vision encoder of the VLM as the data encoder (left and lower part of Fig.~\ref{fig:specializedArchitecture}). Specifically, given $M$ rendered views $X=\{x_i\}_{i=1}^{M}$ of the part model, we extract the embedding from each view and concatenate them to form the multi-view data embeddings:
\begin{equation}
g_i = f_{\mathrm{VE}}(x_i), \quad
G = \operatorname{Concat}\big(\{g_i\}_{i=1}^{M}\big),
\label{eq:data_encoding}
\end{equation}
where $f_{\mathrm{VE}}$ denotes the vision encoder of the VLM. Given the multi-view data embeddings $G$, we further aggregate them into a part-level data embedding to match the dimensionality of the shifted category-level knowledge embeddings. Specifically, mean pooling is used over the view dimension:

\begin{equation}
g \;=\; \operatorname{MeanPool}(G)
\;=\; \frac{1}{M}\sum_{i=1}^{M}g_i.
\label{eq:instance_level_data}
\end{equation}

In the Knowledge–data alignment module, we directly follow Sec.~\ref{sec:method-aligning_knowledge_with_limited_data}. Given the pooled knowledge embedding $\bar{k}_c$ for category $c$, the category-specific shift vector is formulated as:

\begin{equation}
s_c \sim q_{\phi}\!\left(s_c \mid \bar{k}_c, s_c^{(0)}\right),
\quad
s_c^{(0)} =
\frac{1}{|\mathcal{D}_c|}
\sum_{(x_j,y_j)\in \mathcal{D}_c}
g^{(j)},
\label{eq:shift_vector_category}
\end{equation}
where $\mathcal{D}_c=\{(x_j,y_j)\mid y_j=c\}$ denotes the training samples of category $c$, and $g^{(j)}=\operatorname{MeanPool}(G_j)$ denotes the part-level data embedding of sample $x_j$, computed in the same way as Eq.~\eqref{eq:instance_level_data}. Here, $s_c^{(0)}$ is the data embedding center of category $c$, and $s_c$ is further learned during training. Then, the shifted category-level knowledge embeddings are obtained via the addition operator:
\begin{equation}
\tilde{k}_c = \bar{k}_c + s_c.
\label{eq:shifted_knowledge}
\end{equation}

To align the shifted knowledge embeddings with the part-level data embedding $g$, we stack all shifted knowledge embeddings into a memory matrix $\tilde{K}=[\tilde{k}_1;\ldots;\tilde{k}_{|\mathcal{C}|}] \in \mathbb{R}^{|\mathcal{C}|\times d}$ and compute the category-level attention scores with $g$ as the query and $\tilde{K}$ as the key memory:

\begin{equation}
a = \operatorname{MultiHeadAttnScore}(g,\tilde{K})
\label{eq:aligned_data}
\end{equation}
where $a=[a_1,\ldots,a_{|\mathcal{C}|}]$ denotes the category-level attention score vector before softmax, and $d$ is the embedding dimension.

Finally, a classification head is designed to use the attention score vector $a$ for part classification. The attention score vector is normalized by a softmax function to obtain the classification probabilities. It can be formulated as follows:

\begin{equation}
p(c \mid a) =
\frac{\exp(a_c)}
{\sum_{i=1}^{|\mathcal{C}|} \exp(a_i)},
\label{eq:cosine_softmax}
\end{equation}
where $a_c$ denotes the attention score related to category $c$.

The entire knowledge calibration network is trained with the standard cross-entropy loss, while keeping the VLM frozen. The loss is formulated as follows:

\begin{equation}
\mathcal{L}_{\mathrm{CE}}
= - \log p(y \mid a)
= - \sum_{c=1}^{|\mathcal{C}|} \mathbb{I}[y=c]\log p(c \mid a),
\label{eq:cross_entropy}
\end{equation}
where $y$ is the ground-truth label for the part category, and $\mathbb{I}[\cdot]$ is the indicator function.

\begin{figure}[t]
  \centering
  \includegraphics[width=0.48\textwidth]{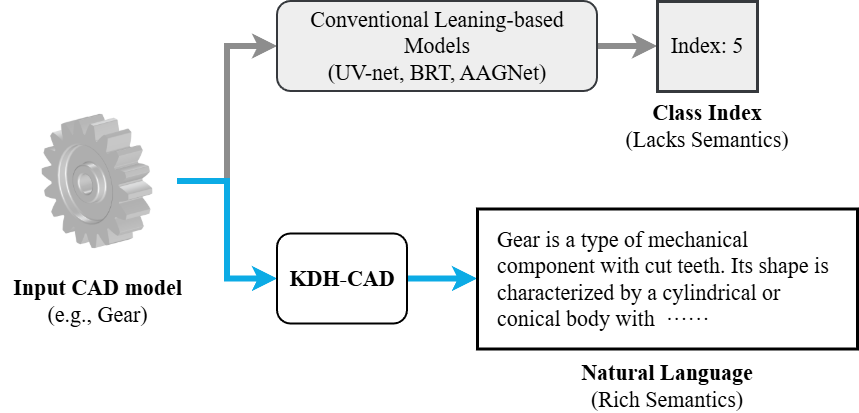}
  \caption{Semantic output using gear as an example.}
  \label{fig:semantic_output_cases}
\end{figure}

It should be noted that our network can also output a semantic description for the query part, as shown in Fig.~\ref{fig:semantic_output_cases}. That is, the semantic description includes the category name, the typical shape, common variants, and the main use, instead of a dumb index. It is an auxiliary output enabled by the VLM. This capability arises because the category-level knowledge embeddings are extracted from the VLM and thus reside in its native representation space. As a result, they can be directly fed back to the VLM as conditioning context for decoding, enabling the output of semantic descriptions. Specifically, given a part, we first obtain the class posterior $p(y \mid a)$ following the above formulations and take the most probable category as the predicted label:

\begin{equation}
\hat{c} \;=\; \arg\max_{c \in \mathcal{C}} \; p(c \mid a).
\label{eq:pred_label}
\end{equation}

The predicted category is then used to retrieve its corresponding token-level knowledge embedding $K_{\hat{c}} \in \mathbb{R}^{L_{\hat{c}}\times d}$ (Eq.~\eqref{eq:token_level_knowledge}), which is fed back into the VLM to generate a textual description for the input part model. Concretely, the generated text sequence $\hat{O}=\{\hat{o}_t\}_{t=1}^{T}$ is obtained by
\begin{equation}
\hat{O} \;=\; \arg\max_{o} \prod_{t=1}^{T}
P_{\mathrm{VLM}}\!\left(o_t \mid o_{<t},\, K_{\hat{c}}\right),
\label{eq:autoregressive_generation}
\end{equation}
where $P_{\mathrm{VLM}}(\cdot)$ denotes the VLM's autoregressive generation probability conditioned on the corresponding category-level knowledge embeddings.

\section{Results and Discussion}
\label{sec:result}

In this section, the proposed KDH-CAD method is evaluated by comparison with baseline models and ablation studies. The experimental setup is described in Sec.~\ref{sec:exp-setup}. Then, learning performance under data scarcity is examined in Sec.~\ref{sec:exp-learning}. Comparisons with existing learning-based methods are presented in Sec.~\ref{sec:exp-comp}, followed by ablation studies in Sec.~\ref{sec:exp-ablation}. Finally, discussions are provided in Sec.~\ref{sec:exp-discussion}.

\subsection{Setup}
\label{sec:exp-setup}

\subsubsection{Datasets}

\textbf{TMCAD Dataset.} The Truly Mechanical CAD (TMCAD) dataset consists of 9,799 authentic mechanical engineering models selected from publicly available online sources. It is organized into 10 categories: \textit{Bearing}, \textit{Bolt-Screw} (now unified), \textit{Bracket}, \textit{Coupling}, \textit{Flange}, \textit{Gear}, \textit{Nut}, \textit{Pulley}, \textit{Spring}, and \textit{Shaft}. The dataset is hosted at: \url{https://github.com/Qiang-Zou/BRT}. More details could be found in the paper~\cite{zou2025bringing}.

\textbf{FabWave Dataset.} The FabWave dataset features 4,504 annotated 3D geometries distributed across 45 distinct categories. Note that while the entire repository encompasses over 100,000 models, category labels are available for a small part of the entries.
For the labeled dataset, we computed the instance count for each category and retained only those categories containing no fewer than 200 models for experimental evaluation. Concretely, \textit{Washers}, \textit{O Rings}, \textit{Bushing}, \textit{Shaft Collar}, \textit{Grommets}, \textit{Keyway Shaft}, \textit{Rotary Shaft}, and \textit{Socket Head Screws}, 8 categories in total, were selected. Moreover, among the 8 categories, Rotary Shaft and Keyway Shaft had an 83\% repetition rate, so they were merged into a single \textit{Shafts} category.

\subsubsection{Training Details}

The foundation model adopted in KDH-CAD is Qwen3-VL-2B-Instruct~\cite{Qwen3-VL}, an off-the-shelf model with approximately 2 billion parameters. It remains frozen throughout the training of our knowledge calibration network. In the knowledge calibration network, the Knowledge–data alignment module uses 8 attention heads for cross-attention. The network is trained using the AdamW optimizer with a weight decay of $1 \times 10^{-4}$ and an initial learning rate of $1 \times 10^{-4}$. A cosine learning rate scheduler is employed to facilitate stable convergence.

To maintain balanced supervision across categories, a per-category split is adopted for TMCAD. Specifically, for each category, the first 100 samples are used for training, the next 50 samples for validation, and the remaining samples for testing. The implementation is based on PyTorch, with FlashAttention2~\cite{dao2023flashattention2} used to accelerate attention computation. Training runs for 40 epochs on a single NVIDIA A100-SXM4-40GB GPU. For evaluation, the checkpoint with the highest validation accuracy is selected and reported on the test set.

\begin{table*}[t]
    \centering
    \small
    \caption{Classification results with 10 runs of random data splits.}
    \label{tab:robustnessProof}
    \setlength{\tabcolsep}{2.5mm}
    \begin{tabular*}{0.96\textwidth}{@{\extracolsep{\fill}} c cccc cccc @{}}
        \toprule
        & \multicolumn{4}{c}{\textbf{TMCAD}} & \multicolumn{4}{c}{\textbf{FabWave}} \\
        \cmidrule(lr){2-5} \cmidrule(lr){6-9}
        \textbf{Run}
        & ACC & BACC & Macro-F1 & Micro-F1
        & ACC & BACC & Macro-F1 & Micro-F1 \\
        \midrule
        Run 1  & 95.43 & 93.54 & 94.09 & 95.43 & 99.84 & 99.75 & 99.82 & 99.84 \\
        Run 2  & 94.64 & 92.04 & 92.73 & 94.64 & 99.84 & 99.75 & 99.84 & 99.84 \\
        Run 3  & 95.17 & 93.55 & 93.71 & 95.17 & 99.95 & 99.86 & 99.90 & 99.95 \\
        Run 4  & 94.10 & 91.60 & 91.85 & 94.10 & 99.95 & 99.86 & 99.90 & 99.95 \\
        Run 5  & 95.50 & 93.67 & 94.01 & 95.50 & 100.00 & 100.00 & 100.00 & 100.00 \\
        Run 6  & 95.57 & 94.01 & 94.36 & 95.57 & 99.52 & 99.19 & 99.44 & 99.52 \\
        Run 7  & 94.54 & 92.77 & 92.98 & 94.54 & 98.93 & 99.54 & 99.20 & 98.93 \\
        Run 8  & 95.42 & 93.90 & 94.22 & 95.42 & 99.62 & 99.60 & 99.58 & 99.62 \\
        Run 9  & 93.44 & 90.58 & 91.31 & 93.44 & 99.46 & 99.28 & 99.46 & 99.46 \\
        Run 10 & 95.15 & 93.13 & 93.48 & 95.15 & 99.68 & 99.69 & 99.63 & 99.68 \\
        \midrule
        \textbf{Mean}
        & \textbf{94.90} & \textbf{92.88} & \textbf{93.27} & \textbf{94.90}
        & \textbf{99.68} & \textbf{99.65} & \textbf{99.68} & \textbf{99.68} \\
        \textbf{Std Dev}
        & \textbf{0.70} & \textbf{1.13} & \textbf{1.04} & \textbf{0.70}
        & \textbf{0.32} & \textbf{0.26} & \textbf{0.26} & \textbf{0.32} \\
        \bottomrule
    \end{tabular*}
\end{table*}

\subsubsection{Evaluation Metrics}
We adopt four standard metrics to evaluate KDH-CAD and other methods. They are defined as follows:

\begin{itemize}
\item \textbf{Accuracy (ACC)}. ACC measures the proportion of correctly classified samples in the test set:
\begin{equation}
\mathrm{ACC}=\frac{1}{N}\sum_{i=1}^{N}\mathbb{I}[\hat{y}_i=y_i],
\end{equation}
where $N$ is the number of test samples, $y_i$ and $\hat{y}_i$ denote the ground-truth and predicted labels of sample $i$, and $\mathbb{I}[\cdot]$ is the indicator function.

\item \textbf{Balanced Accuracy (BACC)}. BACC averages the per-class recall to reduce the influence of class frequency:
\begin{equation}
\mathrm{BACC}=\frac{1}{C}\sum_{c=1}^{C}\frac{\mathrm{TP}_c}{\mathrm{TP}_c+\mathrm{FN}_c},
\end{equation}
where $C$ is the number of classes, and $\mathrm{TP}_c$ and $\mathrm{FN}_c$ are the true positives and false negatives for class $c$.

\item \textbf{Macro-F1}. Macro-F1 computes the F1-score independently for each class and then averages them, treating all classes equally:
\begin{equation}
\mathrm{Macro\text{-}F1}=\frac{1}{C}\sum_{c=1}^{C}\mathrm{F1}_c,
\
\mathrm{F1}_c=\frac{2\,\mathrm{TP}_c}{2\,\mathrm{TP}_c+\mathrm{FP}_c+\mathrm{FN}_c},
\end{equation}
where $C$ is the number of classes, and $\mathrm{TP}_c$, $\mathrm{FP}_c$, and $\mathrm{FN}_c$ denote the numbers of true positives, false positives, and false negatives for class $c$, respectively.

\item \textbf{Micro-F1}. Micro-F1 aggregates the numbers of true positives, false positives, and false negatives over all classes before computing the F1-score, thereby weighting classes by their sample frequencies:
\begin{equation}
\mathrm{Micro\text{-}F1}=\frac{2\sum_{c=1}^{C}\mathrm{TP}_c}{2\sum_{c=1}^{C}\mathrm{TP}_c+\sum_{c=1}^{C}\mathrm{FP}_c+\sum_{c=1}^{C}\mathrm{FN}_c},
\end{equation}
where $C$ is the number of classes, and $\mathrm{TP}_c$, $\mathrm{FP}_c$, and $\mathrm{FN}_c$ denote the numbers of true positives, false positives, and false negatives for class $c$, respectively.

\end{itemize}

\begin{figure}[t]
  \centering
  \includegraphics[width=0.48\textwidth]{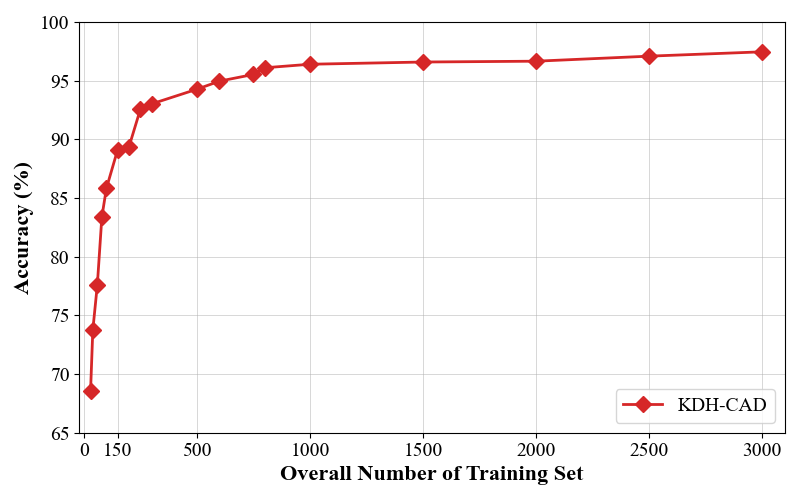}
  \caption{Classification accuracy under varying training dataset sizes (30-3000).}
  \label{fig:accuracyCurve}
\end{figure}

\subsection{KDH-CAD Learning Performance}
\label{sec:exp-learning}

\subsubsection{Learning Accuracy under Varying Training Set Size}
\label{sec:Learning-Accuracy-under-Varying-raining-Set-Size}

To evaluate KDH-CAD learning accuracy under various training dataset sizes, we vary the number of training samples $n$ per category according to the partitioning strategy described in the training details. Specifically, experiments are conducted with $n \in \{3,\allowbreak 4,\allowbreak 5,\allowbreak 6,\allowbreak 8,\allowbreak 10,\allowbreak 15,\allowbreak 20,\allowbreak 25,\allowbreak 30,\allowbreak 50,\allowbreak 60,\allowbreak 75,\allowbreak 80,\allowbreak 100,\allowbreak 150,\allowbreak 200,\allowbreak 250,\allowbreak 300\}$.

As shown in Fig.~\ref{fig:accuracyCurve}, the classification accuracy improves rapidly in the low-data regime, indicating that KDH-CAD can effectively leverage knowledge in foundation models to learn discriminative representations with only limited data. As the training set size increases, performance improvements exhibit diminishing returns, and $n=100$ emerges as a practical trade-off point, achieving strong accuracy with relatively low data. Such a curve suggests that, supported by domain knowledge, the proposed approach can learn high-quality embeddings from limited data, and the embeddings can be effectively utilized for downstream tasks.
\begin{figure*}[t]
  \centering
  \includegraphics[width=0.96\textwidth]{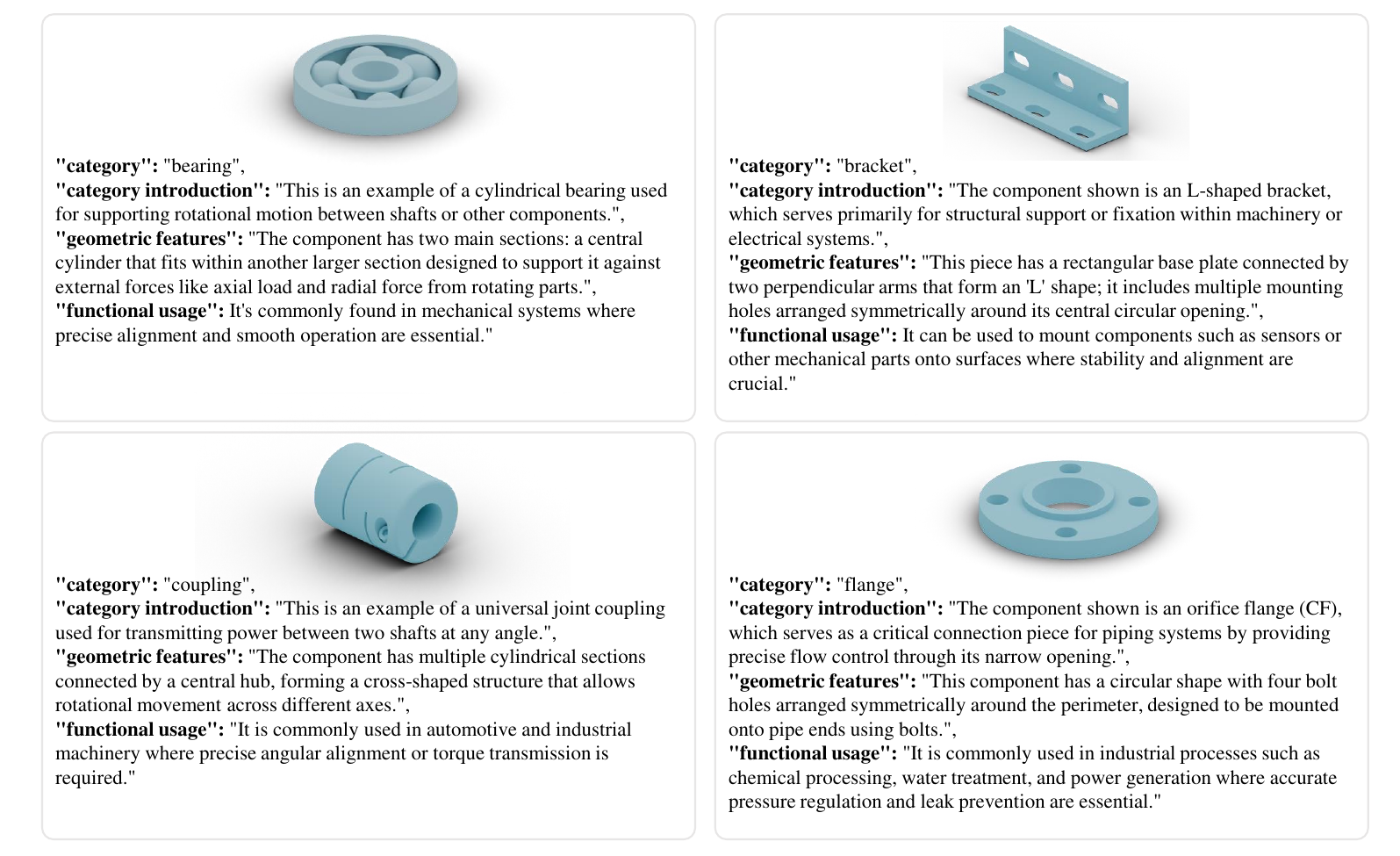}
  \caption{Examples of semantic outputs generated by KDH-CAD.}
  \label{fig:semantic_output_examples}
\end{figure*}
\begin{figure*}[t]
  \centering
  \includegraphics[width=0.96\textwidth]{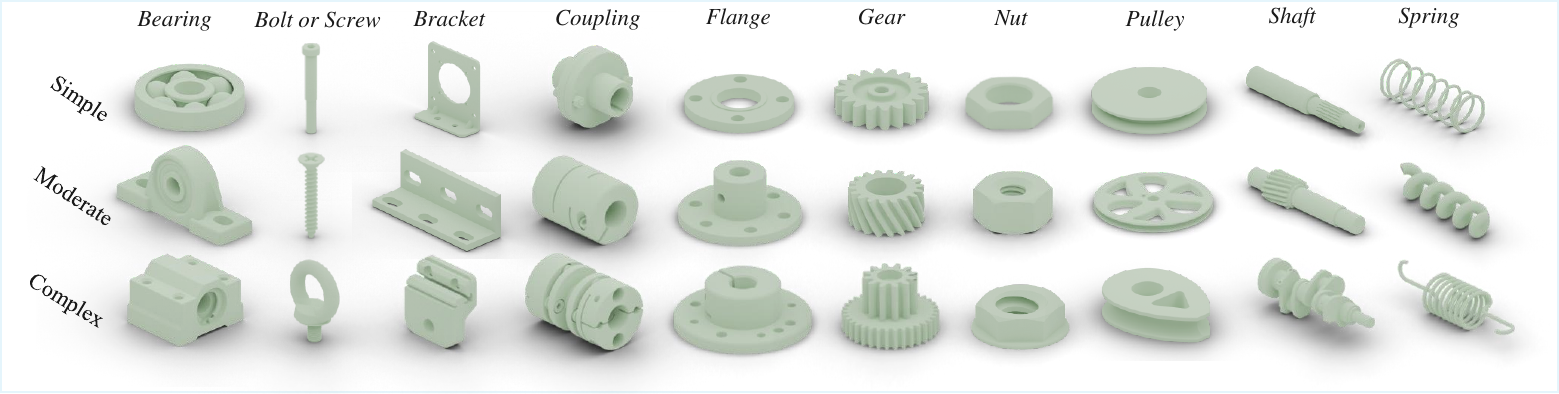}
  \caption{Typical classification results of KDH-CAD.}
  \label{fig:cases}
\end{figure*}

\subsubsection{Training Dataset Robustness}

To assess the stability of KDH-CAD and rule out bias from specific training samples, we conduct a robustness analysis using random data partitioning. With a small and fixed set of training samples, ten independent trials are performed with different random seeds. Specifically, 50 training samples and 50 validation samples per category are randomly selected in each trial for both TMCAD and FabWave, with the remaining samples reserved for testing on TMCAD, 50 samples for testing on FabWave.

The results, summarized in Table~\ref{tab:robustnessProof}, show that KDH-CAD maintains consistent classification accuracy across all random splits, with negligible variance. Since the data split varies across trials, multiple metrics remain stable, and the training loss shows no evident overfitting trend, KDH-CAD appears robust to sample selection and unlikely to be overfitting.

\subsubsection{Semantic Output and Case Studies}

Beyond quantitative evaluations, we briefly show the auxiliary semantic outputs of KDH-CAD, which can provide more interpretable information than traditional index-based category outputs. Specifically, KDH-CAD can generate category-aware textual descriptions, including the part category, characteristic geometry, and typical functional usage. Since existing CAD datasets do not provide standard annotations for part-level semantic descriptions, we evaluate these outputs by expert manual verification. The assessment focuses on semantic correctness, consistency with CAD geometry, and coverage of the main functional and geometric characteristics. Fig.~\ref{fig:semantic_output_examples} shows examples of such outputs. Fig.~\ref{fig:cases} further shows that KDH-CAD correctly classifies CAD models with varying complexity, indicating robustness across different geometric appearances.
\begin{figure}[!t]
  \centering
  \includegraphics[width=0.48\textwidth]{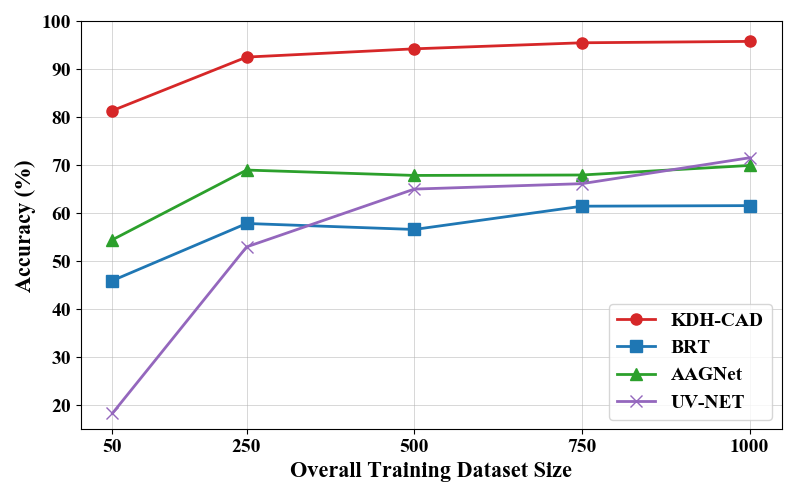}
  \caption{Comparisons of KDH-CAD and existing methods under varying dataset sizes (50--1000).}
  \label{fig:accuracyCurveComparison}
\end{figure}

\subsection{Comparisons}
\label{sec:exp-comp}

Comparisons are conducted on the TMCAD and FabWave datasets, and three representative models are selected as baseline models:
\begin{itemize}
    \item \textbf{UV-Net~\cite{2021_Model_Retrieval_AI_based_UVNet}}: It is a widely cited method that has been adopted as a standard baseline in many prior works;
    \item \textbf{AAGNet~\cite{2024_Wu_AAGNet}}: It proposed new features (e.g., geometric Attributed Adjacency Graph) to perform better than UV-Net on multiple tasks like semantic segmentation, which is widely used as a baseline model in recent CAD learning works;
    \item \textbf{BRT (boundary representation transformer)~\cite{zou2025bringing}}: It is the state-of-the-art approach by now.
\end{itemize}
Since AAGNet is originally designed for segmentation tasks, we adapt it for classification following the setting in the paper~\cite{zou2025bringing}. The other two models were implemented using their official open-source code. In addition to quantitative metrics, we qualitatively compare the model outputs. KDH-CAD has the ability to generate semantic outputs, whereas the baseline models cannot, as shown in Table \ref{tab:qualitative_output_capability}.

\begin{table}[t]
\centering
\small
\caption{Qualitative comparison on output of part classification methods.}
\label{tab:qualitative_output_capability}
\resizebox{0.48\textwidth}{!}{
\begin{tabular}{lcc}
\hline
\textbf{Network} & \textbf{Category Label} & \textbf{Semantic Description} \\
\hline
UV-Net     & Yes & No \\
AAGNet     & Yes & No \\
BRT        & Yes & No \\
\textbf{KDH-CAD (Ours)} & Yes & Yes \\
\hline
\end{tabular}
}
\end{table}

\subsubsection{Comparisons under Data Scarcity}

\begin{table}[!t]
    \centering
    \caption{Quantitative comparisons of classification results. TM indicates the TMCAD dataset, and FW indicates the FabWave dataset.}
    \label{tab:fulldata}
    \footnotesize
    \resizebox{0.48\textwidth}{!}{
    \setlength{\tabcolsep}{6pt}
    \begin{tabular}{ll c cccc}
        \toprule
        \textbf{Network} & \textbf{Dataset} & \textbf{Shots} & \textbf{ACC} & \textbf{BACC} & \textbf{Macro-F1} & \textbf{Micro-F1} \\
        \midrule

        \multirow{2}{*}{UV-Net} & TM & 8,299 & 87.70 & 87.71 & 87.66 & 87.70 \\
                                & FW & 1,962 & 87.50 & 74.89 & 74.35 & 87.50 \\
        \midrule
        \multirow{2}{*}{AAGNet} & TM & 8,299 & 72.00 & 72.00 & 68.25 & 71.98 \\
                                & FW & 1,962 & 85.67 & 85.67 & 88.79 & 85.67 \\
        \midrule
        \multirow{2}{*}{BRT}    & TM & 8,299 & 82.65 & 79.54 & 79.39 & 82.65 \\
                                & FW & 1,962 & 99.68 & 99.40 & 99.55 & 99.68 \\
        \midrule
        \multirow{2}{*}{\textbf{Ours}} & \textbf{TM} & \textbf{1000} & \textbf{95.82} & \textbf{94.50} & \textbf{94.47} & \textbf{95.82} \\
                                       & \textbf{FW} & \textbf{350}  & \textbf{99.84} & \textbf{99.75} & \textbf{99.82} & \textbf{99.84} \\
        \bottomrule
    \end{tabular}
    }
\end{table}

We compare KDH-CAD with UV-Net, BRT, and AAGNet on the TMCAD and FabWave datasets under data scarcity. Specifically, we vary the number of training samples per category from 5 to 100. For fair comparison, all models are trained using the same training-validation-test split and the same number of epochs.

As shown in Fig.~\ref{fig:accuracyCurveComparison}, when the number of training samples per category varies from 5 to 100, KDH-CAD consistently outperforms the baselines, particularly in low-data regimes.

Moreover, while purely data-driven models experience substantial performance degradation as the training size decreases, KDH-CAD remains stable. This robustness arises from its ability to leverage cross-modal alignment and inherit semantic priors from the foundation model, enabling effective generalization even with minimal supervision.

\begin{figure}[t]
  \centering
  \includegraphics[width=0.48\textwidth]{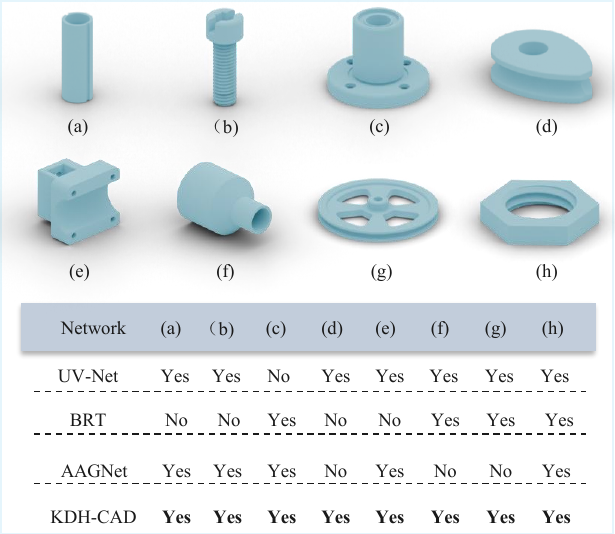}
  \caption{Part classification examples by KDH-CAD and existing methods. Each row indicates whether the model in the first column correctly predicted the result, e.g., part (d) is mispredicted only by UV-Net.}
  \label{fig:casesFullDataComparison}
\end{figure}

\begin{figure*}[b]
  \centering
  \includegraphics[width=0.96\textwidth]{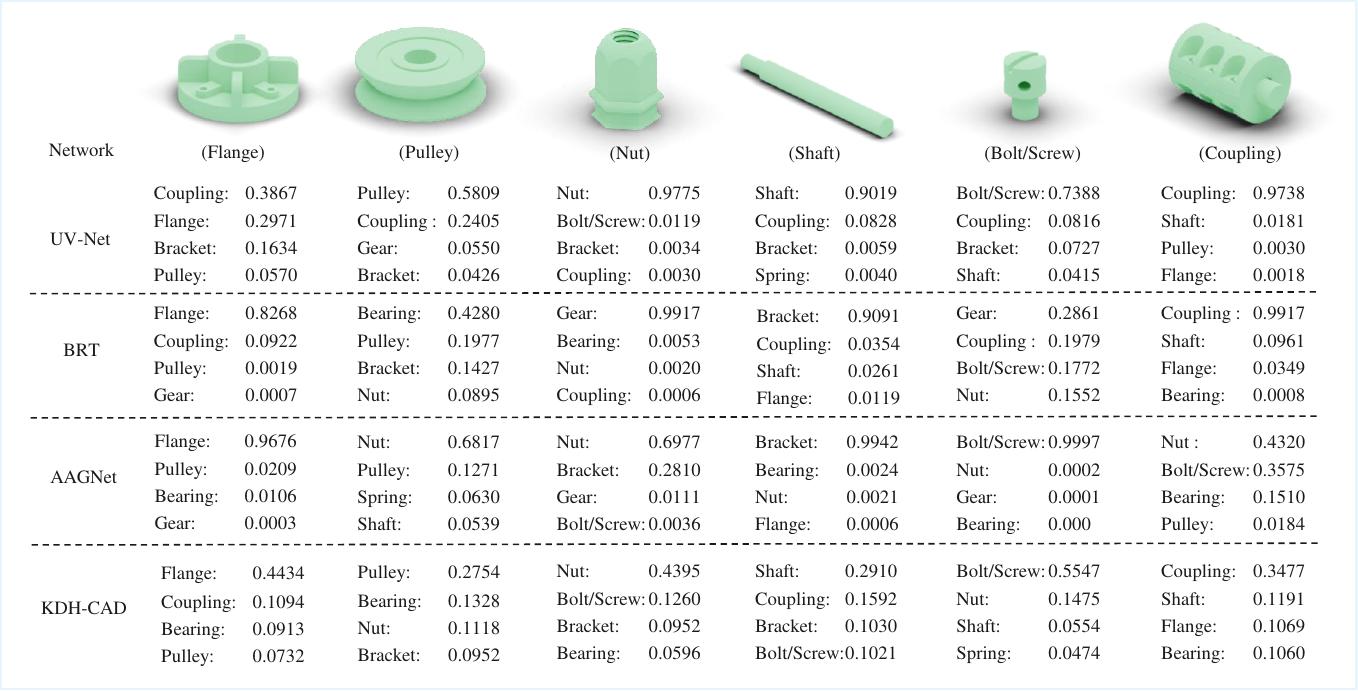}
  \caption{Comparisons of the top-4 probabilities of part classification on typical CAD models. Each column corresponds to a CAD part model shown above; the name in parentheses indicates the ground-truth label. For each row, the top-4 predicted category names and probabilities are listed in descending order.}
  \label{fig:case_probability}
\end{figure*}

\begin{figure*}[t]
  \centering
  \includegraphics[width=0.96\textwidth]{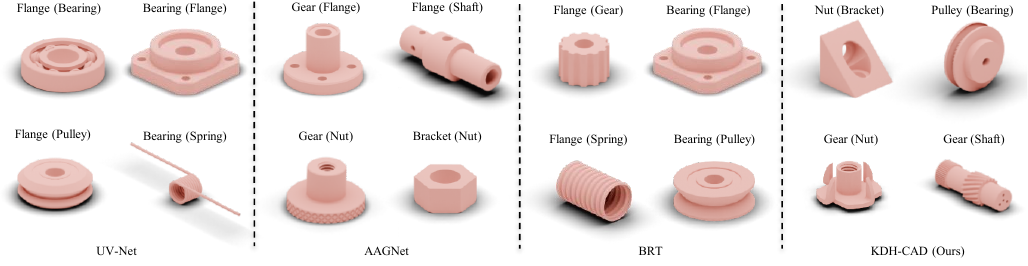}
  \caption{Typical failed examples of existing methods and KDH-CAD. Flange (Bearing) means the AI model mistakenly classifies a bearing as a flange, and the same for others.}
  \label{fig:failed_cases_of_KDHCAD}
\end{figure*}

\subsubsection{Comparisons under Full Data}
\label{sec:exp-Comparisons-under-Full-Data}

We further compare KDH-CAD with UV-Net, BRT, and AAGNet on the TMCAD and FabWave datasets in a relatively unfair way. Specifically, the baselines are trained on all samples in TMCAD and FabWave datasets (8,299 and 1,962, respectively), whereas KDH-CAD is trained on a subset with only 100 and 50 samples per category, respectively. For all methods, identical validation and test splits are used: the final 50 samples of each category for validation and the preceding 50 for testing. This setup highlights the gap between Knowledge–data hybrid learning and purely data-driven approaches.

As shown in Table~\ref{tab:fulldata}, KDH-CAD achieves higher accuracy under limited training data, indicating that its performance gains stem from the integration of knowledge and data. Although more training data improves the baselines, KDH-CAD remains highly data-efficient. As shown in Fig.~\ref{fig:accuracyCurve}, it approaches the strongest baseline trained with about 10k samples using only about 100 samples, and exceeds most baseline metrics with about 150 samples. The results further suggest that leveraging external knowledge can substantially reduce the dependence on large-scale CAD data, which is important for practical CAD applications where such datasets are often difficult to obtain.

Fig.~\ref{fig:casesFullDataComparison} illustrates representative part classification results on the TMCAD dataset in comparison with three other methods. KDH-CAD correctly predicts all the shown samples, whereas the other models fail on one or more cases. Furthermore, Fig.~\ref{fig:case_probability} shows the top-4 probabilities for six additional parts. Although KDH-CAD produces lower top-1 confidence in some cases in Fig.~\ref{fig:case_probability}, it still makes correct predictions and achieves higher accuracy than the baselines. Since confidence can be affected by softmax or logit scaling and may be overestimated for incorrect predictions~\cite{zhu2022rethinking}, prediction correctness should be regarded as the primary performance indicator.

\subsubsection{Comparisons on Model Retrieval}
\label{sec:exp-model-retrieval}

To further evaluate the practical value of KDH-CAD, we add a CAD model retrieval experiment using the same data split as Sec.~\ref{sec:exp-Comparisons-under-Full-Data}. Retrieval is performed with cosine similarity, and evaluated by Recall@$K$ and mAP. The retrieval representation is taken before the final classification layer for all methods.

As shown in Table~\ref{tab:cad_retrieval}, KDH-CAD achieves the best results on all metrics, including 96.18\% Recall@1 and 96.85\% mAP. Although trained with substantially fewer samples, it still outperforms UV-Net by 9.89\% in Recall@1 and 20.14\% in mAP, demonstrating its practical value for CAD model retrieval.

\begin{table*}[t]

\centering
\caption{Performance comparison of CAD model retrieval.}
\label{tab:cad_retrieval}
\small
\renewcommand{\arraystretch}{1.15}
\setlength{\tabcolsep}{6pt}

\begin{tabular*}{0.96\textwidth}{@{\extracolsep{\fill}}lccccc}
\toprule
Model & Training dataset size & Recall@1 & Recall@5 & Recall@10 & mAP \\
\midrule
AAGNet  & 10,000 & 75.20\% & 79.03\% & 81.65\% & 70.48\% \\
BRT     & 10,000 & 80.41\% & 87.28\% & 90.84\% & 73.18\% \\
UV-Net  & 10,000 & 86.29\% & 91.94\% & 93.55\% & 76.71\% \\
KDH-CAD & 1,000  & \textbf{96.18\%} & \textbf{97.47\%} & \textbf{97.66\%} & \textbf{96.85\%} \\
\bottomrule
\end{tabular*}
\end{table*}

\begin{table}[!t]

\centering
\caption{Statistics of running time and memory usage.}
\label{tab:runtime_memory}
\footnotesize
\resizebox{0.48\textwidth}{!}{
\setlength{\tabcolsep}{6pt}
\begin{tabular}{lcc}
\toprule
Model  & Inference Speed & GPU Memory Usage  \\
\midrule
AAGNet   & 126.13 ms/sample  & 34.5 MB   \\
UV-Net      & 20.18 ms/sample  & 310.2 MB   \\
BRT   & 47.45 ms/sample & 317.9 MB  \\
KDH-CAD  & 319.52 ms/sample & 2.02 GB  \\
\bottomrule
\end{tabular}
}
\end{table}

\begin{table*}[t]
    \centering
    \small
    \caption{Ablation studies of KDH-CAD.}
    \label{tab:ablation_combined}
    \renewcommand{\arraystretch}{1.2}

    \begin{tabular*}{0.96\textwidth}{@{\extracolsep{\fill}}lcccc}
        \hline
        \textbf{Configuration} & \textbf{ACC} & \textbf{BACC} & \textbf{Macro-F1} & \textbf{Micro-F1} \\
        \hline
        W/O Domain Knowledge or Data & 83.81 & 84.07 & 81.83 & 83.81 \\
        W/O Domain Knowledge         & 87.82 & 86.91 & 85.53 & 87.82 \\
        W/O Data                     & 86.29 & 81.77 & 73.51 & 86.29 \\
        W/O Cross-Attention or Shift & 11.58 & 10.29  & 2.61  & 11.58 \\
        W/O Cross-Attention          & 86.94 & 86.20 & 84.23 & 86.94 \\
        W/O Shift                    & 94.72 & 93.53 & 93.49 & 94.72 \\
        \hline
        \textbf{KDH-CAD (Full)}      & \textbf{95.82} & \textbf{94.50} & \textbf{94.47} & \textbf{95.82} \\
        \hline
    \end{tabular*}
\end{table*}

\subsubsection{Running Time and Memory Usage}

To provide a more complete comparison, we evaluate inference time and GPU memory usage on the same NVIDIA A100-SXM4-40GB GPU. The average per-sample inference time and GPU memory usage are measured on the same 500 test samples, with results reported in Table~\ref{tab:runtime_memory}.

As shown in Table~\ref{tab:runtime_memory}, KDH-CAD requires more computation than the baselines, mainly due to the use of a foundation model. Nevertheless, it processes one sample in 319.52 ms and uses 2.02 GB of GPU memory on average, about 5\% of the A100 40GB GPU. Since part classification and model retrieval are mainly offline CAD tasks, this overhead remains practical given the substantial performance improvement under limited labeled data.

\subsection{Ablation Studies}
\label{sec:exp-ablation}
We carry out ablation experiments by altering or removing individual components of KDH-CAD on TMCAD dataset. The ablation study is structured as follows:

\begin{enumerate}[label=(\alph*)]
    \item \textbf{KDH-CAD:} The KDH-CAD's network, incorporating all modules and knowledge sources.
    \item \textbf{Without domain knowledge and data:} Direct inference using the pretrained Qwen3-VL-2B-Instruct model with simple prompts.
    \item \textbf{Without domain knowledge:} Linear probing on data embeddings without leveraging domain knowledge.
    \item \textbf{Without data:} Direct inference with domain knowledge as prompts, without parameter updates.
    \item \textbf{Without cross-attention and shift:} The cross-attention fusion module is removed, and the shift vector is disabled during training, no parameter updates.
    \item \textbf{Without cross-attention:} The cross-attention module is removed, while the shift vector mechanism is retained.
    \item \textbf{Without shift:} The cross-attention fusion module is preserved, but the shift vector is disabled.
\end{enumerate}

The ablation results on TMCAD are presented in Table~\ref{tab:ablation_combined}. As shown in the ablation studies, removing domain knowledge or data degrades performance, and removing both further limits the model to the frozen VLM prior, indicating that limited data lacks explicit CAD semantic priors while knowledge alone cannot adapt to the target geometric distribution. In addition, the alignment module is critical. Without both cross-attention and shift, cosine similarity directly compares data embeddings with knowledge embeddings in the original VLM latent space. Since this space is mainly optimized for generation and does not naturally provide well-aligned multimodal embeddings~\cite{jiang2025vlm2vec}, this causes a severe modality gap and leads to the 11.58\% accuracy. The smaller drops from removing either module show that both contribute to alignment. The cross-attention aligns features, while the shift vector adapts knowledge embeddings to data distribution.

\subsection{Discussion}
\label{sec:exp-discussion}

As illustrated in Fig.~\ref{fig:accuracyCurve}, KDH-CAD is seen to give a meaningful performance gain under data scarcity, reaching $\sim$95\% accuracy on the part classification task with only 75 training samples per category, 750 samples in total. This indicates that our method is data efficient. Moreover, Table~\ref{tab:robustnessProof} shows that KDH-CAD achieves consistently high accuracy across multiple random subsamplings under the same low-data regime. It suggests that, in the evaluated part classification setting, our method is robust to the noise and variability inherent in small training sets.

Comparisons with existing CAD learning methods were carried out on both the TMCAD and FabWave datasets. As shown in Fig.~\ref{fig:accuracyCurveComparison}, KDH-CAD achieves consistently higher accuracy across a range of low-data settings, with an improvement of approximately 20 percentage points over the compared methods. Notably, KDH-CAD remains the best method even when the other methods are trained with substantially more labeled data (e.g., 5--8$\times$ larger training sets), as shown in Table~\ref{tab:fulldata}. This is likely due to our method leveraging CAD knowledge as a prior, thereby reducing reliance on learning task-relevant regularities solely from data.

It can also be observed that failure cases exist in KDH-CAD, as illustrated in Fig.~\ref{fig:failed_cases_of_KDHCAD}. These failures occur more frequently for parts that are very complex or relatively uncommon. This is likely because the mechanical design textbooks and tutorial videos used as our domain knowledge primarily describe canonical structures and typical design patterns, and thus do not comprehensively cover application-specific variants and special cases. For such parts, accurate classification may require specialized, application-dependent knowledge beyond what is covered in general instructional materials. However, application-dependent knowledge is often proprietary and resides within companies in the form of internal documentation, design standards, and accumulated design examples.
Future work may therefore incorporate such proprietary knowledge to adapt KDH-CAD to specialized industrial tasks and improve its applicability. Additionally, the rendered multi-view images used in the current implementation cannot fully preserve the exact CAD topology and geometry of the original models. However, the core idea of KDH-CAD is not inherently limited to image input, and future work can extend it to native CAD inputs, for example by incorporating B-rep encoders such as UV-Net.

\section{Conclusion}
\label{sec:conclusion}

In this paper, we treat CAD learning under data scarcity as a knowledge completion and calibration problem, and introduce KDH-CAD, a Knowledge–data hybrid learning framework to reduce the reliance on large-scale CAD data.
Specifically, KDH-CAD leverages the latent space of the foundation model as a transferable representation space. Within this latent space, structured domain knowledge is translated into knowledge embeddings that capture elicited and completed CAD-relevant concepts. Then, a small amount of labeled CAD data is utilized to calibrate these embeddings to solve the concept-geometry mismatch.
This method is instantiated on the part classification task and demonstrates state-of-the-art performance under data scarcity on real-world mechanical part classification datasets. The current implementation is limited to part classification and classification-related model retrieval, and broader validation on other CAD learning tasks remains future work.

Despite KDH-CAD's strong performance in low-data regimes, several limitations have been observed in our experiments. Most notably, the method may fail on complex or relatively uncommon parts. A key reason is that the domain knowledge used in this work is intentionally generic and mainly covers canonical structures and typical patterns, leaving less support for specialized variants. Correctly recognizing these parts may therefore require more application-dependent knowledge. In practice, this knowledge is often proprietary and remains within companies as internal resources, which are difficult to share or centralize due to confidentiality constraints. A promising direction is therefore to extend KDH-CAD with a federated learning~\cite{mcmahan2017communication} setting, enabling companies to collaboratively improve the model while keeping their private resources local. This may provide a practical pathway to enhance KDH-CAD’s applicability to specialized variants and complex cases.

\section*{Acknowledgements}

This work has been funded by the ``Pioneer" and ``Leading Goose" R\&D Program of Zhejiang Province (No. 2024C01103), the National Natural Science Foundation of China (No. 62102355), and the Fundamental Research Funds for the Zhejiang Provincial Universities (No. K20250142, K20241957).

\section*{References}

\bibliography{mybibfile}

% ===== Appendices imported from supplementary_clean.tex =====
\appendix

\section{Construction of Limited Data}

\begin{table*}[htbp]
\centering
\small
\caption{Number of models per part category in the TMCAD dataset.}
\label{tab:tmcad_counts}
\renewcommand{\arraystretch}{1.2}

\begin{tabular*}{0.96\textwidth}{@{\extracolsep{\fill}}lccccccccccc}
\toprule
Category & Bearing & Bolt-Screw & Bracket & Flange & Gear & Nut & Shaft & Coupling & Pulley & Spring & Total\\
\midrule
Number of Parts & 857 & 2,070 & 1,102 & 979 & 962 & 899 & 893 & 474 & 544 & 1,019 & 9,799\\
\bottomrule
\end{tabular*}
\end{table*}

The dataset used in our experiments is the Truly Mechanical CAD (TMCAD) dataset~[20], which comprises ten mechanical part categories: \textit{bearing}, \textit{bolt or screw}, \textit{bracket}, \textit{coupling}, \textit{flange}, \textit{gear}, \textit{nut}, \textit{pulley}, \textit{shaft}, and \textit{spring}. The number of samples per category is summarized in Table~\ref{tab:tmcad_counts}. The CAD models are collected from open-access online repositories, with each model annotated by human with a category name label. All parts are provided in STEP format. To further ensure data quality, the entire dataset undergoes an additional manual verification and de-duplication process, guaranteeing both label accuracy and file integrity. Owing to their high similarity in geometric characteristics and functional roles, the originally separate \textit{bolt} and \textit{screw} categories are merged into a single class in our experiments, while the original fine-grained labels are retained for potential future use. The dataset and its download link are publicly available at \url{https://github.com/Qiang-Zou/BRT}.

\begin{figure*}[t]
    \centering
    \includegraphics[width=1.0\textwidth]{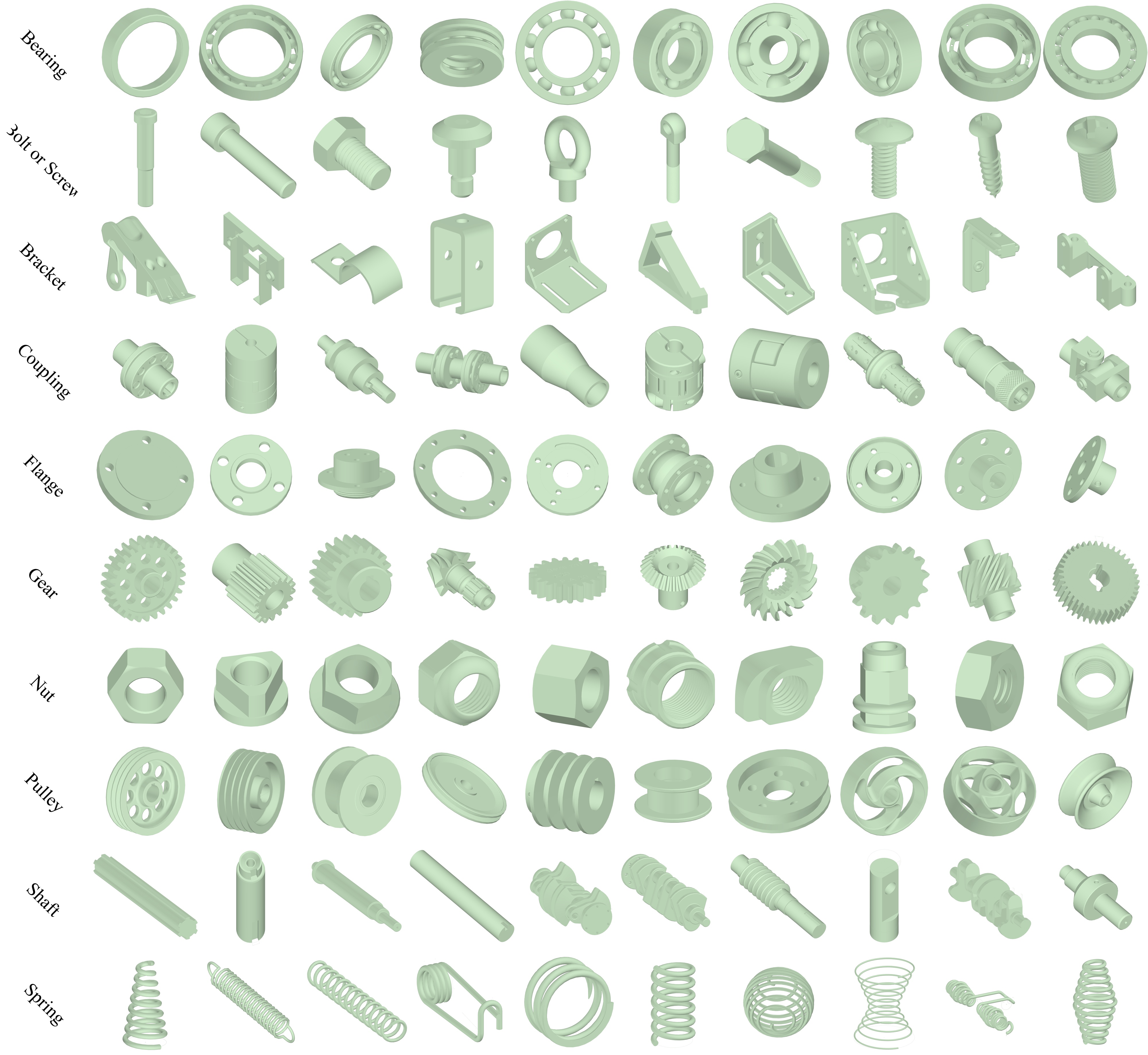}
    \caption{Overview of the TMCAD dataset. Each row represents one of the ten mechanical part categories, with ten representative instances per row showcasing geometric diversity.}
    \label{fig:dataset_overview}
\end{figure*}

To simulate limited-data learning scenarios, we restrict the amount of training data to construct constrained training sets based on the TMCAD and FabWave datasets. Fig.~\ref{fig:dataset_overview} presents ten representative instances for each mechanical part category in TMCAD. For TMCAD, two data split strategies are adopted.

\textbf{Strategy I: Index-based split.}
Samples are ordered according to their original index in the dataset. This strategy is used in Sections~4.2.1, 4.3.1, 4.3.2, and~4.4, with different training set sizes depending on the specific experiment.
For the experiments in Section~4.2.1, the first $n$ samples from each category are used for training, where $n \in \{3, 4, 5, 6, 8, 10, 15, 20, 25, 30, 50, 60, 75, 80, 100, 150, 200, 250, 300\}$.
For the experiments in Section~4.3.1, the first $n$ samples from each category are used for training, where $n \in \{5, 25, 50, 75, 100\}$. The subsequent 50 samples are used for validation, and the remaining samples are used for testing.
For the experiments in Section~4.3.2, the first 100 samples from each category are used for training. The last 100 samples are reserved for evaluation, with the first 50 used for validation and the remaining 50 for testing.
For the experiments in Section~4.4, the first 100 samples from each category are used for training, followed by 50 samples per category for validation, while the remaining 8,299 samples are reserved for testing.
Since the dataset was randomly shuffled during annotation and verification, this index-based split does not correlate with geometric complexity, data source, or category-specific attributes.

\textbf{Strategy II: Random split.}
This strategy is used in Section~4.2.2. Ten random splits are carried out, where 50 samples per category are selected for training and 50 samples per category are selected for validation, with the remaining samples used for testing. The random seed ranges from 10 to 100 with a step size of 10.

\begin{table*}[htbp]
\centering
\small
\caption{Number of models per part category in the FabWave dataset (selected categories).}
\label{tab:fabwave_counts}
\renewcommand{\arraystretch}{1.2}

\begin{tabular*}{0.96\textwidth}{@{\extracolsep{\fill}}lcccccccc}
\toprule
Category &
Bushing &
Grommets &
Shafts &
O Rings &
Shaft Collar &
Socket Head Screws &
Washers &
Total \\
\midrule
Number of Parts &
310 & 269 & 500 & 376 & 281 & 204 & 722 & 2,662\\
\bottomrule
\end{tabular*}
\end{table*}

With respect to the FabWave dataset, only a limited portion of it is annotated with category labels, and only eight categories contain more than 200 parts. Therefore, we select these categories to balance the dataset and to provide sufficient training data for all compared methods, including \textit{bushing}, \textit{grommets}, \textit{shafts}, \textit{O-rings}, \textit{shaft collar}, \textit{socket head screws}, and \textit{washers}. Table~\ref{tab:fabwave_counts} reports the number of parts in each selected category. Furthermore, we observe that approximately 83\% of the samples are duplicated between the \textit{Rotary Shaft} and \textit{Keyway Shaft} categories; consequently, we merge them into a single \textit{shafts} category to avoid ambiguity during training.

In addition, the FabWave dataset used in our experiments also adopts two data split strategies.
\textbf{(i)} In Section~4.2.2, 50 samples are randomly selected for training, 50 samples are used for validation, and the remaining samples are reserved for testing. The random seed ranges from 10 to 100 with a step size of 10.
\textbf{(ii)} In Section~4.3.2, to ensure consistent validation and test sets across all methods, we first shuffle the entire dataset using a fixed random seed (42). For the baseline models, samples indexed from 50 to 100 are used for validation, samples from 100 to 150 are used for testing, and all remaining samples are used for training. In contrast, for KDH-CAD, only the first 50 samples are selected for training, whereas the validation and test sets remain the same as those used for the baseline models, making the KDH-CAD training set a strict subset of the baseline training data.

\section{Engineering Knowledge Triplet}

To construct the engineering knowledge triplets, we first collected one representative introductory webpage and one demonstration video for each mechanical part category from publicly available online sources. These collected materials serve as the source domain knowledge for each part category. To make the collected webpages compatible with the vision--language model used in our experiment (Qwen3-VL-2B-Instruct~[9]), we further processed them using Trafilatura, a Python-based web content extraction tool. Specifically, each webpage was cleaned and converted into an XML format, from which all textual paragraphs and images within the main content region were extracted while preserving their original relative ordering.

The extracted textual and visual segments, together with the corresponding demonstration video, were then provided to Qwen3-VL only for formatting and normalization. Specifically, Qwen3-VL reorganized the given source materials into an engineering knowledge triplet, consisting of (i) a canonical terminology, (ii) a textual--visual description, and (iii) a visual demonstration, rather than generating new domain knowledge. The model was instructed not to introduce information beyond the provided materials, and the resulting triplets were manually checked before being used in the experiments. During this process, the original image resolutions were preserved, while the video input was uniformly sampled into 12 frames using the Qwen3-VL interface. In addition, we employed the prompt shown in Table~\ref{tab:prompt_ekt} to guide the model in reorganizing all provided resources. Table~\ref{tab:triplets_part1} summarizes the constructed triplets for each part category.

\begin{table*}[htbp]
\centering
\caption{Prompt for Engineering Knowledge Triplet.}
\label{tab:prompt_ekt}
\begin{tabular}{|p{0.9\textwidth}|}
\hline

\begin{minipage}[t]{0.9\textwidth}
\vspace{0.5em}
You are a domain expert in \textbf{mechanical engineering and mechanical part taxonomy}.
Given the provided \textbf{demonstration videos}, \textbf{tutorial webpages}, and \textbf{images} of the webpage, your task is to \textbf{systematically extract, normalize, and reorganize Engineering knowledge Triplet} for the mechanical part class: \textbf{\{class\_name\}}.

\vspace{0.5em}
Focus exclusively on \textbf{engineering semantics} that are essential for understanding, identifying, and categorizing this part class.
Discard any information that is irrelevant, overly generic, marketing-oriented, repetitive, or unrelated to the mechanical function or geometry of the part.

\vspace{0.5em}
The extracted information should be reorganized into the following structured fields, with each field serving a distinct purpose:

\begin{itemize}[leftmargin=1.5em, topsep=2pt, itemsep=2pt]
    \item \textbf{Category}:
    Determine a \textbf{canonical and standardized terminology} for this part class.
    Prefer widely accepted mechanical engineering terms over colloquial or ambiguous names, and avoid dataset-specific or informal aliases, only respond with one term that is the most relevant to the part concept.

    \item \textbf{Description}:
    Provide a comprehensive description covering:
    \begin{itemize}[leftmargin=1.2em, topsep=1pt, itemsep=1pt]
        \item The \textbf{primary mechanical functions} and typical roles of this part in assemblies or systems.
        \item Common \textbf{application scenarios} in machinery, industrial equipment, or mechanical structures.
        \item Key \textbf{geometric characteristics and visual identifiers}, such as typical shapes, structural features, or distinctive components that enable visual recognition.
    \end{itemize}

    \item \textbf{Visual demonstration}:
    Collect and preserve \textbf{all relevant visual evidence} associated with this part class.
    This includes every \texttt{[Image Path]} that depicts the part itself, its structure, or its usage context.
    Do \textbf{not} introduce new images, modify paths, or omit any image that is relevant to the part class.
\end{itemize}

\vspace{0.5em}
\textbf{Output constraints:}
\begin{itemize}[leftmargin=1.5em, topsep=2pt, itemsep=2pt]
    \item Strictly output a single \textbf{JSON object}.
    \item The JSON must contain \textbf{exactly three keys}: \texttt{Category}, \texttt{Description}, and \texttt{Visual demonstration}.
    \item Do not include any additional explanations, comments, or formatting outside the JSON.
    \item Do not extensively omit the original materials.
\end{itemize}
\vspace{0.5em}
\end{minipage}
\\ \hline
\end{tabular}
\end{table*}

\begin{table*}[t]
\centering
\small
\renewcommand{\arraystretch}{1.15}
\setlength{\tabcolsep}{4pt}

\begin{tabularx}{\textwidth}{
>{\centering\arraybackslash}m{1.8cm}
X
>{\centering\arraybackslash}m{4.0cm}
>{\centering\arraybackslash}m{3.4cm}
}

\toprule
\textbf{Canonical Terminology} &
\textbf{Description} &
\textbf{Image} &
\textbf{Visual Demonstration} \\
\midrule

Bearing &
Supporting rotational or linear motion, transmitting loads, and reducing friction and wear between moving parts.
Cylindrical or spherical forms, including ball bearings, roller bearings, plain bearings \ldots &
\begin{adjustbox}{max width=3.8cm, center}
\begin{tabular}{@{}c c c@{}}
\includegraphics[height=1.8cm]{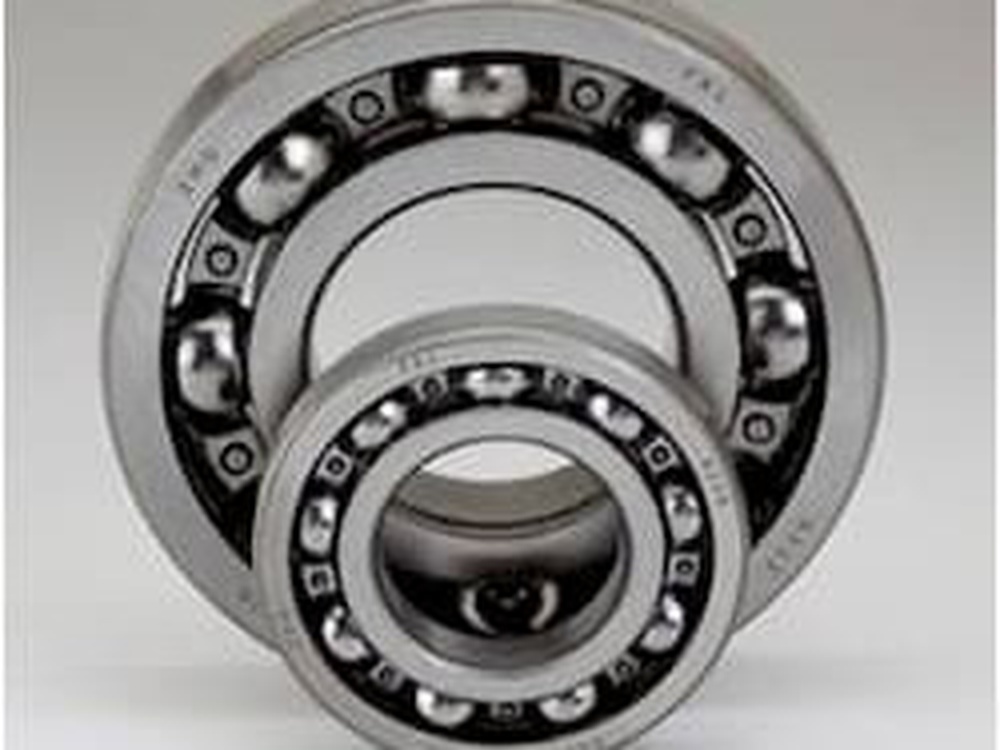} &
\includegraphics[height=1.8cm]{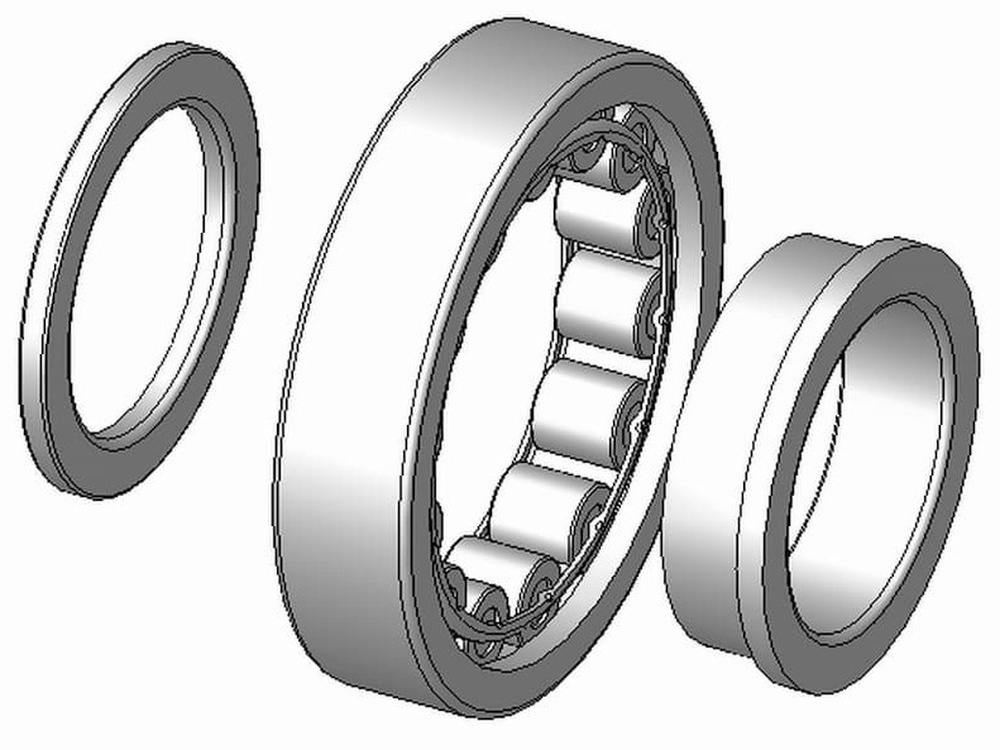}  &
\makebox[0.3cm][c]{\Large$\cdots$}
\end{tabular}
\end{adjustbox} &
\url{https://pan.zju.edu.cn/share/f124228843815fb99a8d8c2624} \\ \addlinespace[4pt]
Bolt / Screw &
Fastening components together by converting rotational motion into axial clamping force.
Threaded shafts with various head types, including hexagonal, flat, oval, and \ldots &
\begin{adjustbox}{max width=3.8cm, center}
\begin{tabular}{@{}c c c@{}}
\includegraphics[height=1.8cm]{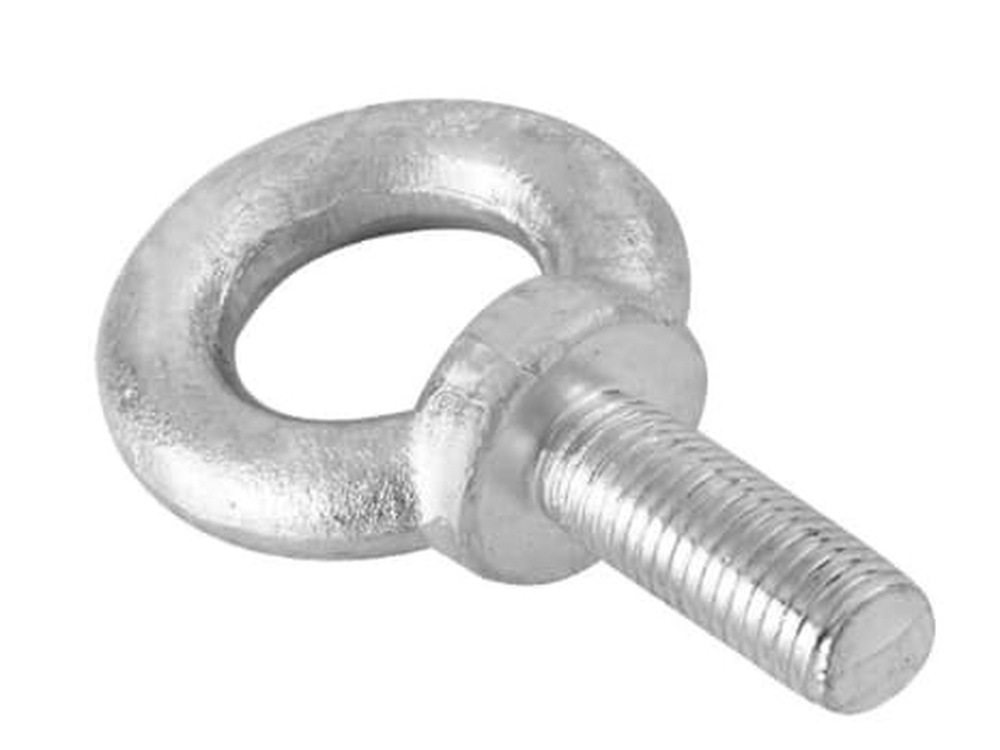} &
\includegraphics[height=1.8cm]{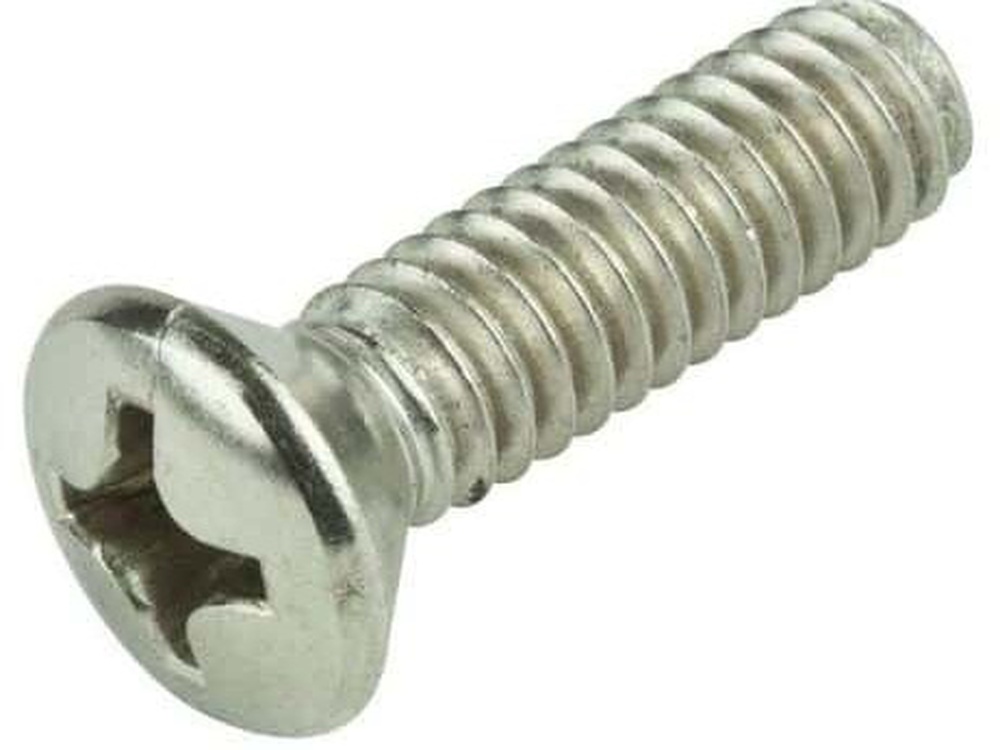} &
\makebox[0.3cm][c]{\Large$\cdots$}
\end{tabular}
\end{adjustbox} &
\url{https://pan.zju.edu.cn/share/a1adc5a34b7f35afcf34789e0d} \\ \addlinespace[4pt]

Bracket &
Providing structural support, reinforcement, and fixation for mechanical or electrical components.
L-shaped, triangular, square, I-shaped, A-shaped, Z-shaped, or \ldots &
\begin{adjustbox}{max width=3.8cm, center}
\begin{tabular}{@{}c c c@{}}
\includegraphics[height=1.8cm]{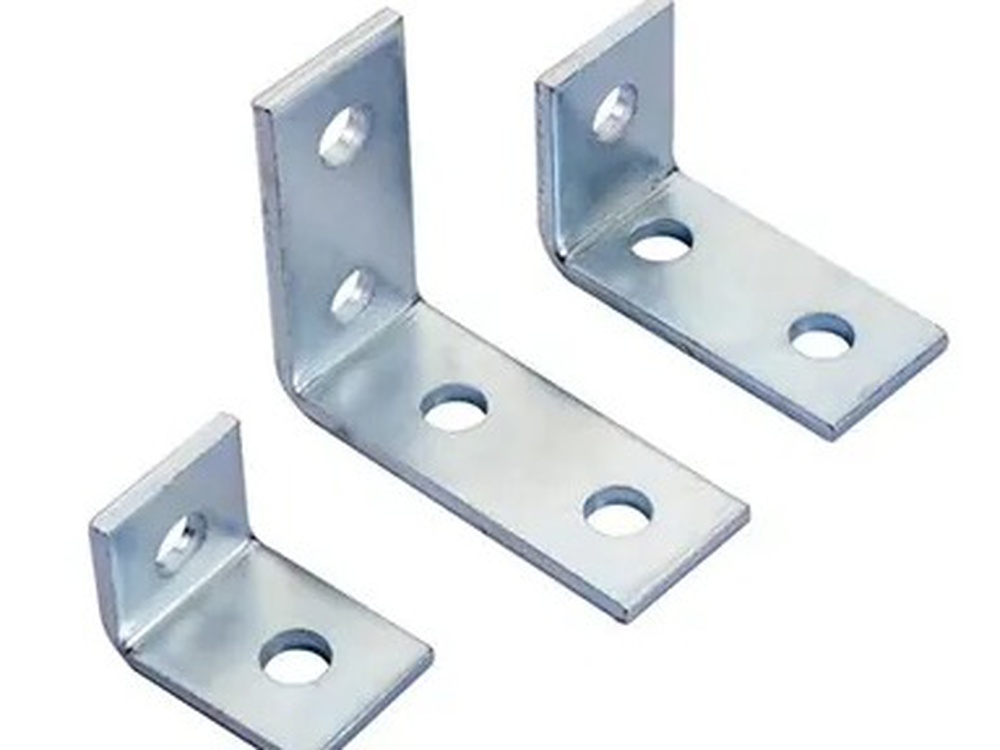} &
\includegraphics[height=1.8cm]{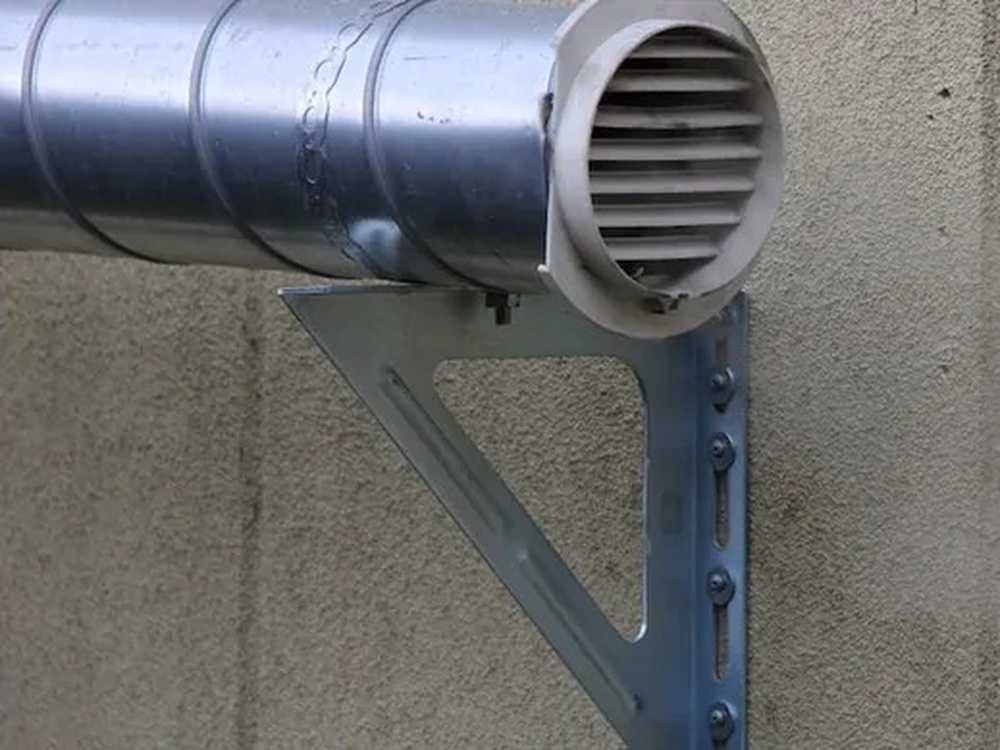} &
\makebox[0.3cm][c]{\Large$\cdots$}
\end{tabular}
\end{adjustbox} &
\url{https://pan.zju.edu.cn/share/1cf87c41067ee5f8376bb6c3c0} \\ \addlinespace[4pt]

Coupling &
Connecting two shafts to transmit power while compensating for misalignment and absorbing vibration.
Flange couplings, gear couplings, rigid and flexible couplings \ldots &
\begin{adjustbox}{max width=3.8cm, center}
\begin{tabular}{@{}c c c@{}}
\includegraphics[height=1.8cm]{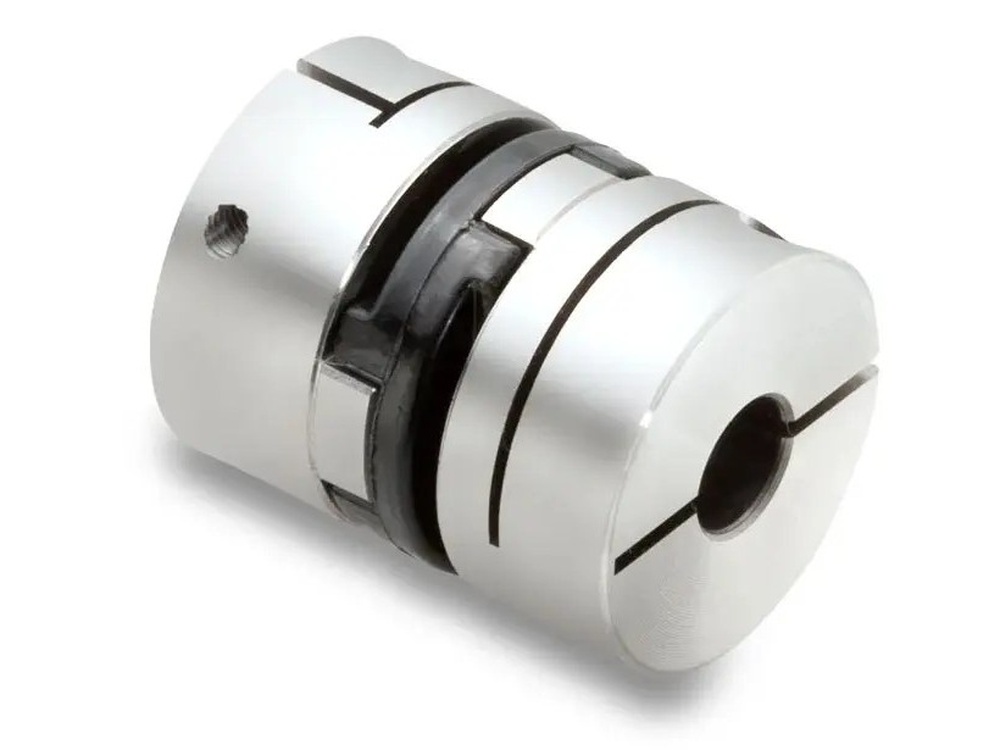} &
\includegraphics[height=1.8cm]{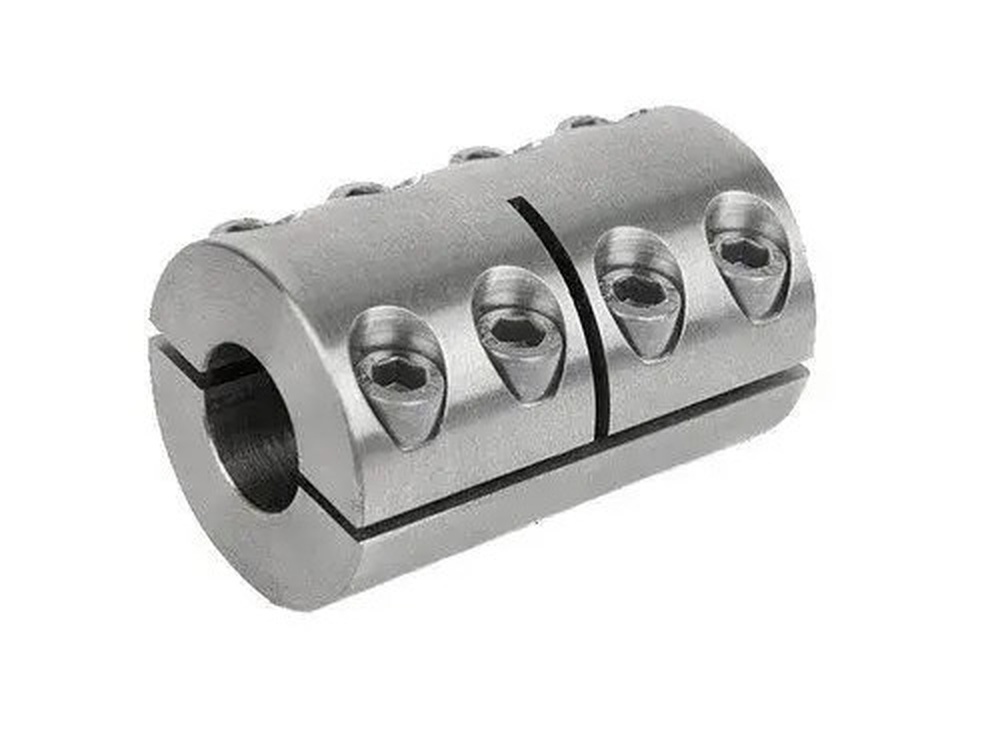} &
\makebox[0.3cm][c]{\Large$\cdots$}
\end{tabular}
\end{adjustbox} &
\url{https://pan.zju.edu.cn/share/9dcad1d6e8c04d5584bc4d8e00} \\ \addlinespace[4pt]

Flange &
Connecting pipes, valves, and equipment in piping systems while enabling inspection and maintenance.
Disc-shaped components with bolt holes, including weld neck \ldots &
\begin{adjustbox}{max width=3.8cm, center}
\begin{tabular}{@{}c c c@{}}
\includegraphics[height=1.8cm]{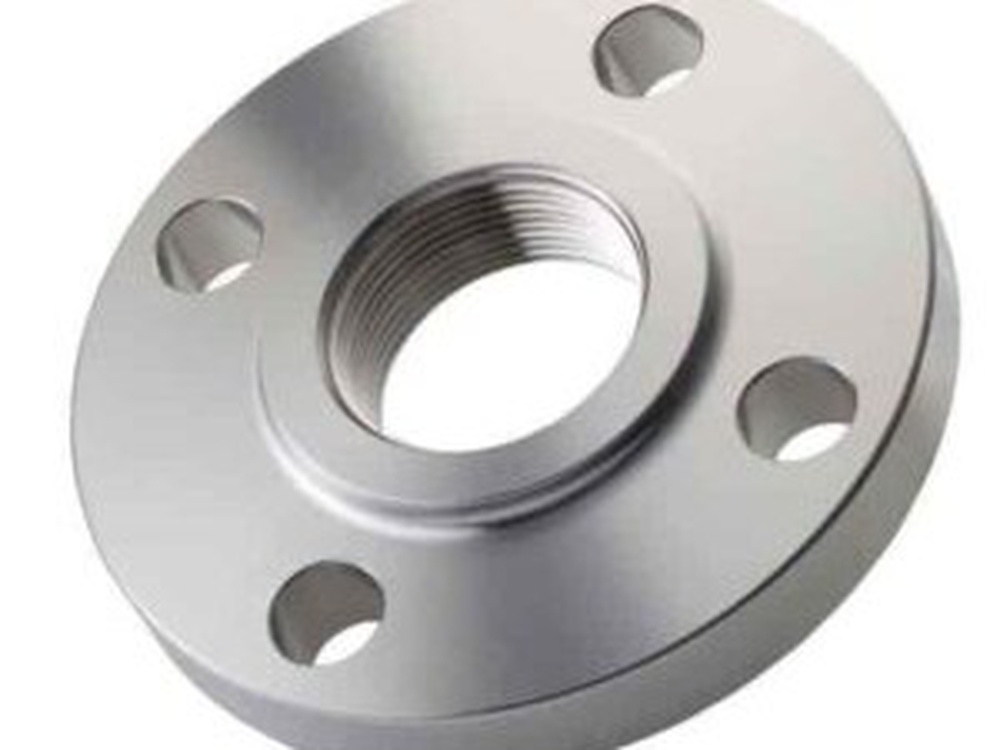} &
\includegraphics[height=1.8cm]{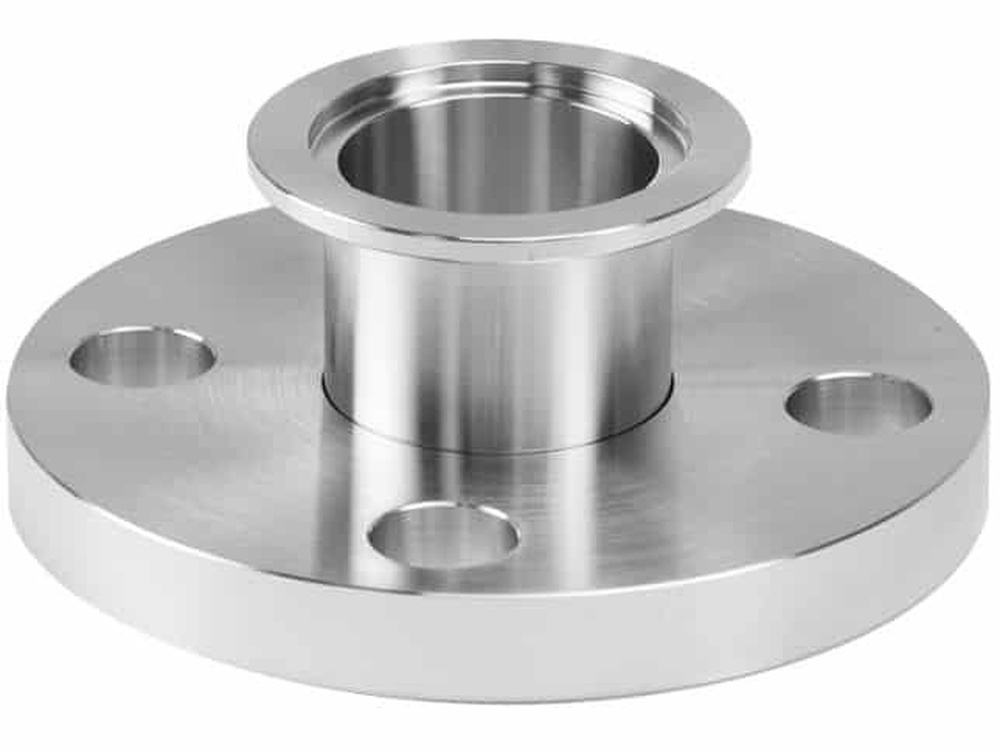} &
\makebox[0.3cm][c]{\Large$\cdots$}
\end{tabular}
\end{adjustbox} &
\url{https://pan.zju.edu.cn/share/663ecfa6c007efefc177badd13} \\ \addlinespace[4pt]

Gear &
Transmitting power, changing speed, torque, or direction of motion in mechanical systems.
Spur, helical, bevel, and worm gears \ldots &
\begin{adjustbox}{max width=3.8cm, center}
\begin{tabular}{@{}c c c@{}}
\includegraphics[height=1.8cm]{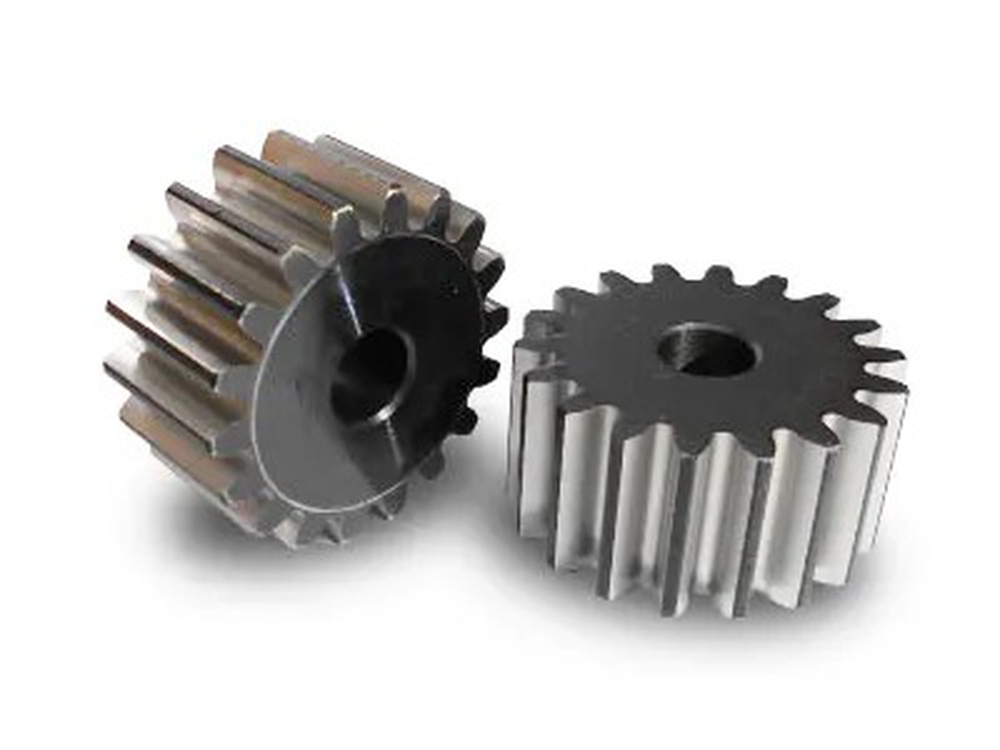} &
\includegraphics[height=1.8cm]{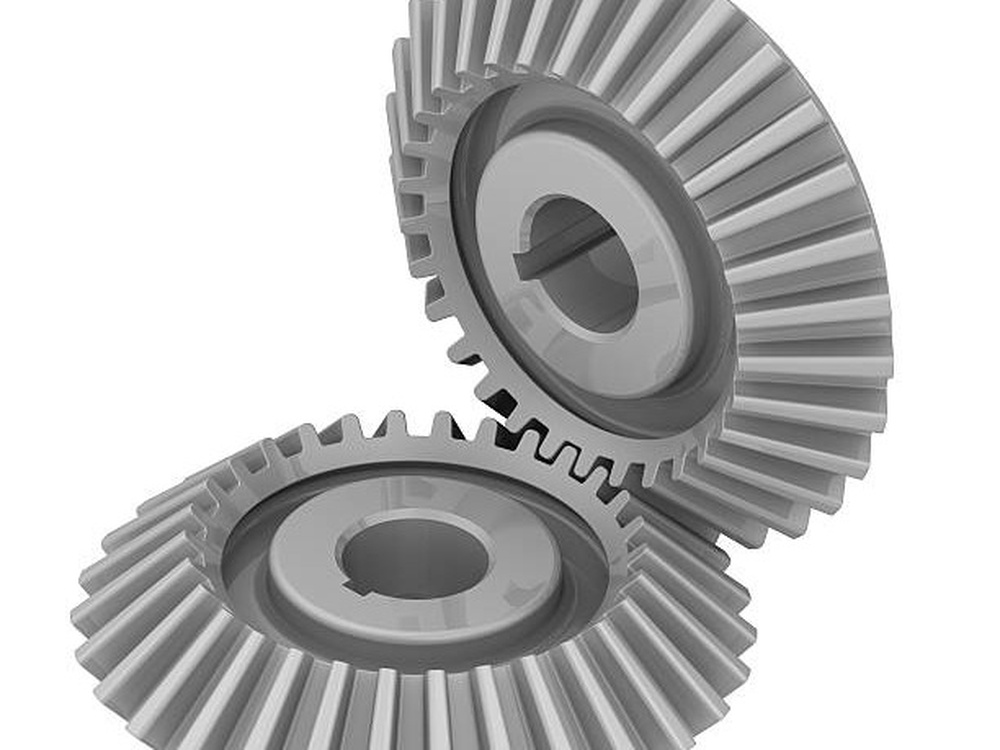} &
\makebox[0.3cm][c]{\Large$\cdots$}
\end{tabular}
\end{adjustbox} &
\url{https://pan.zju.edu.cn/share/8fd8d1030a0da37a7d240ded13} \\ \addlinespace[4pt]

Nut &
Providing threaded fastening by mating with bolts or screws to secure components.
Hex, square, flange, acorn, coupling, cage \ldots &
\begin{adjustbox}{max width=3.8cm, center}
\begin{tabular}{@{}c c c@{}}
\includegraphics[height=1.8cm]{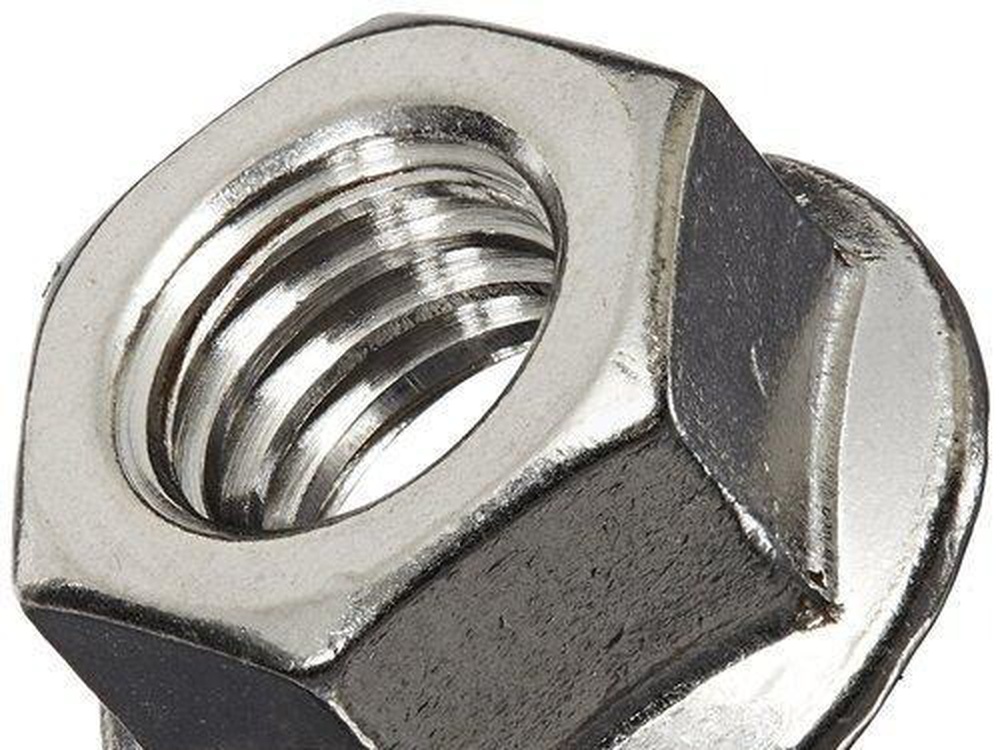} &
\includegraphics[height=1.8cm]{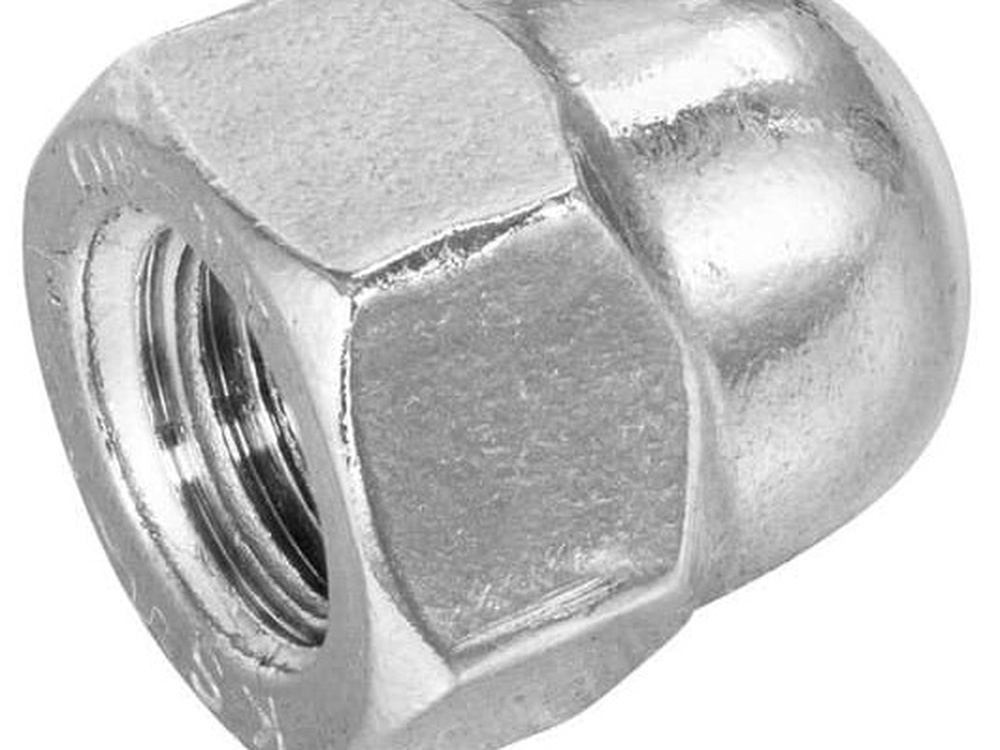} &
\makebox[0.3cm][c]{\Large$\cdots$}
\end{tabular}
\end{adjustbox} &
\url{https://pan.zju.edu.cn/share/1f00409c4250bf7da154a8abd6} \\ \addlinespace[4pt]

Pulley &
Redirecting force and transmitting power via belts, ropes, or chains in lifting and transmission systems.
A wheel with a grooved rim \ldots &
\begin{adjustbox}{max width=3.8cm, center}
\begin{tabular}{@{}c c c@{}}
\includegraphics[height=1.8cm]{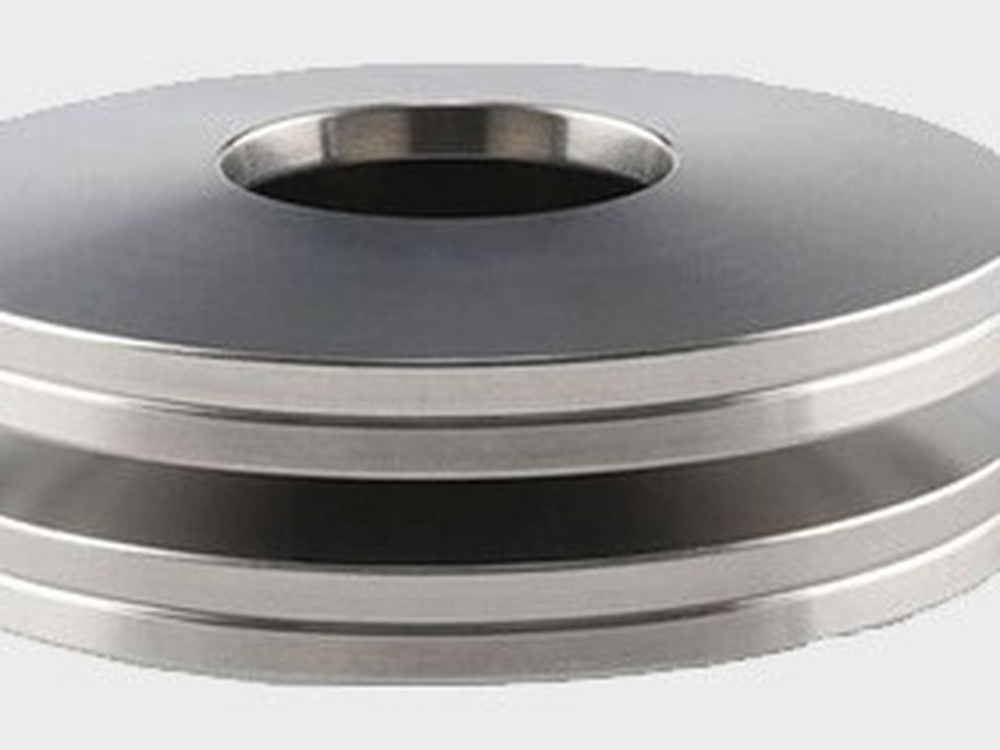} &
\includegraphics[height=1.8cm]{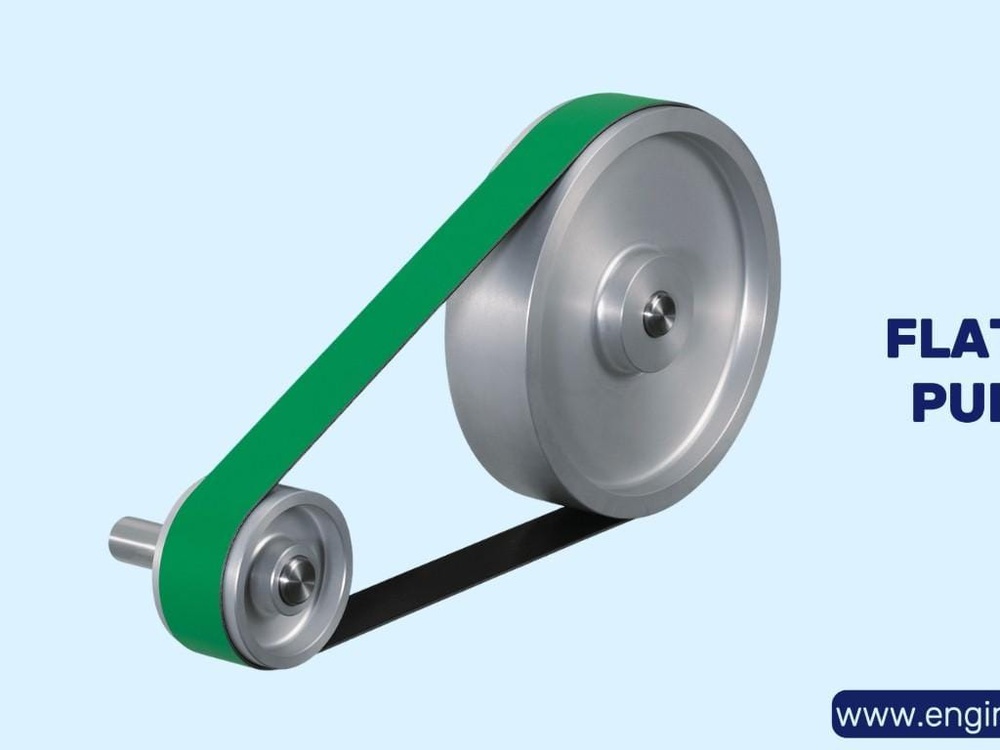} &
\makebox[0.3cm][c]{\Large$\cdots$}
\end{tabular}
\end{adjustbox} &
\url{https://pan.zju.edu.cn/share/053ebdf9668f53e344e201c3bd} \\ \addlinespace[4pt]

Shaft &
Transmitting power and motion between machine components while providing structural support.
Cylindrical or annular cross-sections \ldots &
\begin{adjustbox}{max width=3.8cm, center}
\begin{tabular}{@{}c c c@{}}
\includegraphics[height=1.8cm]{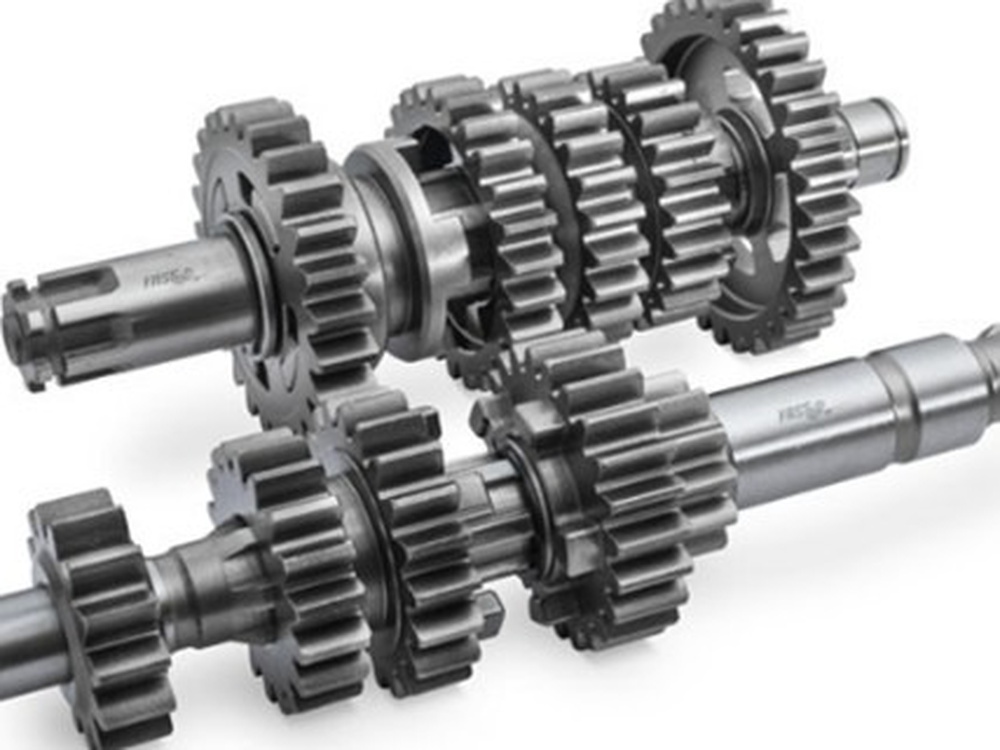} &
\includegraphics[height=1.8cm]{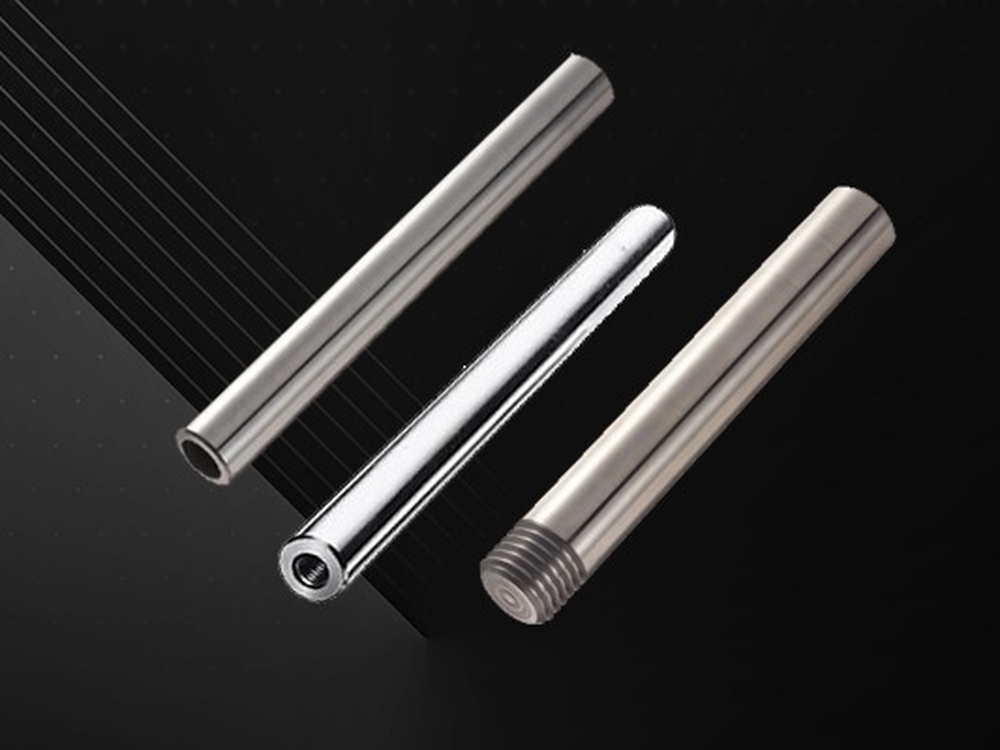} &
\makebox[0.3cm][c]{\Large$\cdots$}
\end{tabular}
\end{adjustbox} &
\url{https://pan.zju.edu.cn/share/3d91277bec1fe5ede31b5f4b1b} \\ \addlinespace[4pt]

Spring &
Storing and releasing mechanical energy, maintaining force, and absorbing shock or vibration.
Helical compression and extension springs \ldots &
\begin{adjustbox}{max width=3.8cm, center}
\begin{tabular}{@{}c c c@{}}
\includegraphics[height=1.8cm]{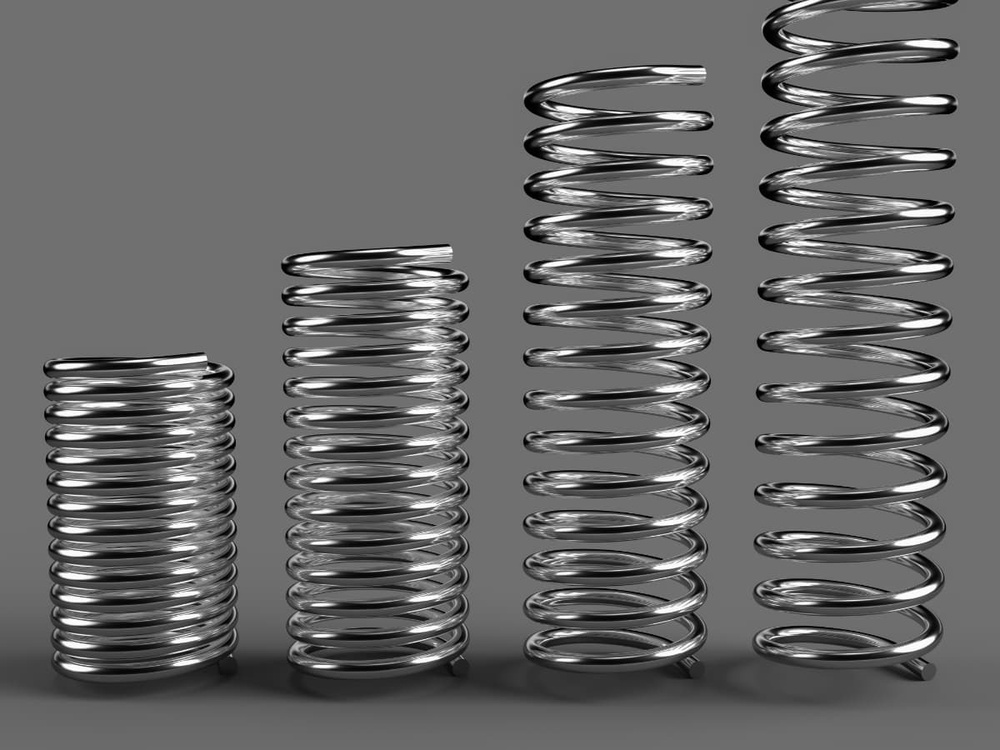} &
\includegraphics[height=1.8cm]{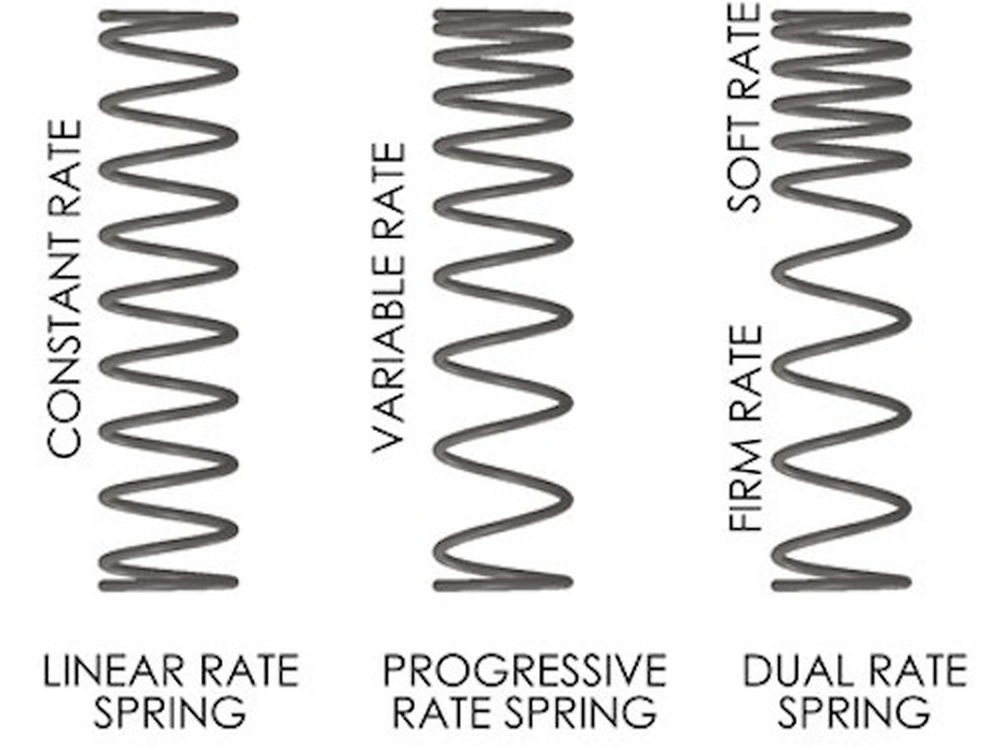} &
\makebox[0.3cm][c]{\Large$\cdots$}
\end{tabular}
\end{adjustbox} &
\url{https://pan.zju.edu.cn/share/b30172bb824f5062766fc84660} \\

\bottomrule
\end{tabularx}

\caption{Engineering Knowledge Triplet for mechanical part categories used in TMCAD.}
\label{tab:triplets_part1}
\end{table*}
% ===== End of imported appendices =====

\end{document}